\documentclass[12pt]{article}
\usepackage[utf8]{inputenc}
\usepackage{amsmath,amsfonts,epsfig,fp}
\def\be{\begin{equation}\begin{gathered}}
\def\ee{\end{gathered}\end{equation}}
\def\Sum{\sum\limits}
\def\half{{\textstyle{1\over2}}}
\def\Prod{\prod\limits}
\def\d{\partial}
\def\pd{\partial}

\def\tr{\mathrm{\,tr\,}}
\def\bs{\boldsymbol}

\def\mc{\mathcal}
\def\Oint{\oint\limits}

\def\Tr{{\rm Tr}}
\def\Int{\int\limits}

\def\bpm{\begin{pmatrix}}
\def\epm{\end{pmatrix}}
\newcommand{\BOXES}[3]{\multiput(#1,#2)(10,0){#3}{\put(0,0){\line(1,0){10}\line(0,1){10}}
\put(10,10){\line(-1,0){10}\line(0,-1){10}}}}
\newcommand{\rf}[1]{(\ref{#1})}

\newcommand{\eq}[1]{\begin{equation}\begin{gathered}#1\end{gathered}\end{equation}}
\newcommand{\eqs}[1]{\begin{equation}\begin{gathered}\begin{split}#1\end{split}\end{gathered}\end{equation}}

\numberwithin{equation}{section}
\newtheorem{theorem}{Theorem}
\newtheorem{lemma}{Lemma}

\title{Free fermions, W-algebras and isomonodromic deformations}

\author{P. Gavrylenko$^{a,b}$, A. Marshakov$^{c,a}$}
\date{\small
$^a${\it Department of Mathematics and Laboratory\\ of Mathematical
Physics, NRU HSE, Moscow, Russia}\\
$^b${\it Bogolyubov Institute for Theoretical Physics, Kyiv, Ukraine}\\
$^c${\it Theory Department, Lebedev Physics Institute and\\
Institute for Theoretical and Experimental Physics, Moscow, Russia}\\
\vspace{0.2cm}
gavrylenko@bitp.kiev.ua,\ \ \ mars@itep.ru\\
\vspace{0.3cm}
\emph{Dedicated to 75-th birthday of Igor Tyutin, contribution \\ to special volume of
Theoretical and Mathematical Physics}
}

\begin{document}

\maketitle

\begin{abstract}
We consider the theory of multicomponent free massless fermions in two dimensions and use it for construction
of representations of W-algebras at integer Virasoro central charges. We define the vertex operators
in this theory in terms of solutions of the corresponding isomonodromy problem. We use this construction to get
some new insights on tau-functions of the multicomponent Toda type hierarchies for the class of solutions, given by the isomonodromy vertex operators and get useful representation for the tau-function of isomonodromic
deformations.
\end{abstract}

\section{Introduction}

The aim of the paper is to present briefly the main free-fermionic constructions that appear in the study of correspondence between the problem of isomonodromic deformations and two-dimensional conformal field theories -- for some class of the theories with extended conformal symmetry.
An interest to the two-dimensional conformal field theories (CFT) with extended nonlinear symmetries, generated by the higher spin holomorphic currents, has been initiated by pioneering work \cite{ZamW}. These theories with so called W-symmetry possess many features of ordinary CFT, including the free field representation \cite{FZ,FL}, which becomes especially simple for the case of integer Virasoro central charges. However, even in this relatively simple case
it turns already to be impossible to construct in generic situation the W-conformal blocks \cite{Wblocks},
which are the main ingredients of the conformal bootstrap definition of the physical correlation functions \cite{BPZ}.

This interest has been seriously supported already in our century by rather nontrivial correspondence between two-dimensional CFT and
four-dimensional supersymmetric gauge theory \cite{LMN,NO,AGT}, where the conformal blocks have to be compared with the Nekrasov instanton partition functions \cite{Nek} producing
in the quasiclassical limit the Seiberg-Witten prepotentials \cite{SW}. This correspondence also meets serious problems beyond $SU(2)$/Virasoro
level: both on four-dimensional gauge theory and two-dimensional CFT sides.
These difficulties can be attacked using different approaches, for example in \cite{GMtw} we have demonstrated
how the exact conformal blocks for the twist fields \cite{ZamAT} in theories with W-symmetry can be computed, using the technique developed previously in \cite{KriW,AMtau,Quiver}.

Here we present another approach to the study of the CFT vertex operators in the theories with extended conformal symmetry, based on their free-fermionic construction. It is clear, that it should work (at least) in the cases of integral central charges, where it is intimately related with the recently discovered there CFT/isomonodromy correspondence \cite{Painleve,PG}.
We are going to discuss the operator content of these theories with nontrivial monodromy properties, and then turn to the problem of computation of the matrix elements of generic monodromy operators. Finally, we are going to relate these matrix elements with the tau-functions of two different classes of problems --  the tau-functions of the multicomponent classical integrable hierarchies of Toda type, and the tau-functions of the isomonodromic deformations.

We are happy to contribute this paper to a special volume, devoted to the 75-th birth of Igor Tyutin.
We know Igor as the best expert in many issues of quantum field theory, and we would like to point out,
that the material of his lecture course ``Conformal field theory from quantum field theory'' has many intersections with
the ideas, developed in this paper. One of the authors also remembers Igor as the most active participant of the special
seminar on Integrable systems at Lebedev Institute, where in particular the fermionic construction of the tau-functions
was intensively discussed. We hope therefore to get some response from Igor to the fermionic constructions,
discussed below.

\section{Abelian $U(1)$ theory
\label{ss:u1}}

\subsection{Fermions and vertex operators}

Introduce the standard two-dimensional holomorphic fermionic fields with the
action $S = {1\over\pi}\int_\Sigma d^2z \tilde{\psi}\d\psi$, so that
\be
\tilde{\psi}(z)\psi(z') = {1\over z-z'} +\ldots
\ee
or
\be
\label{1ferm}
\{\psi_r,\tilde{\psi}_s\} = \delta_{r+s,0},\ \ \ r,s\in \mathbb{Z}+\half\,,
\\
\psi(z)=\Sum_{r\in \mathbb{Z}+\frac12}\frac{\psi_{r}}{z^{r+1/2}}\,,
\hspace{1cm}\tilde{\psi}(z)=\Sum_{s\in \mathbb{Z}+\frac12}
\frac{\tilde{\psi}_s}{z^{s+1/2}}
\ee
with the half-integer mode expansion. The bosonization formulas read
\be
\tilde{\psi}(z)=:e^{i\phi(z)}: = e^{-\Sum_{n<0}\frac{J_n}n z^{-n}}e^{-\Sum_{n>0}\frac{J_n}n z^{-n}}e^ Q z^{J_0}\,,\\
\psi(z)=:e^{-i\phi(z)}: =e^{\Sum_{n<0}\frac{J_n}n z^{-n}}e^{\Sum_{n>0}\frac{J_n}n z^{-n}}e^{- Q} z^{-J_0}\,,\label{bosonization_}
\ee
where
\be
J(z) = :\tilde{\psi}(z)\psi(z): = i\d\phi(z) = \sum_{n\in \mathbb{Z}}{J_n\over z^{n+1}}\,,
\\
[J_n,J_m]=n\delta_{n+m,0},\ \ \ n,m\in \mathbb{Z},\ \ \ \ [J_n, Q]=\delta_{n0}\,,
\label{cur}
\ee
where normal ordering means, that all negative modes stand to the left of all positive, and all $Q$ to the left
of $J_0$.

Consider now generic vertex operators for the bosonic fields
\be
V_{\nu}(z)=:e^{i\nu\phi(z)}: =
e^{-\nu\Sum_{n<0}\frac{J_n}n z^{-n}}e^{-\nu\Sum_{n>0}\frac{J_n}n z^{-n}}e^{\nu Q} z^{\nu J_0} \equiv V_{\nu}^{-}(z)V_{\nu}^{+}(z)e^{\nu Q} z^{\nu J_0}
\label{vertex_}
\ee
which satisfy the obvious exchange relations, following from the Campbell-Hausdorff formula
\be
V_{\alpha}^{+}(z)V_{\beta}^{-}(w) = \left(1-{w\over z}\right)^{\alpha\beta}V_{\beta}^{-}(w)V_{\alpha}^{+}(z)\,,
\\
V_{\alpha}(z)V_{\beta}(w) = \left({z\over w}\right)^{\alpha\beta}\left(1-{w\over z}\right)^{\alpha\beta}\left(1-{z\over w}\right)^{-\alpha\beta}V_{\beta}(w)V_{\alpha}(z)\,.
\ee
One can also write
\be
\label{no}
V_\alpha(z)V_\beta(w) = (z-w)^{\alpha\beta}:V_\alpha(z)V_\beta(w):\,.
\ee
Since vertex operators contain the factor $e^{\nu Q}$, they shift the vacuum charge
\eq{
V_\nu(z):\mc H^{\sigma}\to\mc H^{\sigma+\nu}
}
when acting onto a sector in full Hilbert space
\eq{
\mc H=\bigoplus_\sigma \mc{H}^\sigma
}
corresponding to the definite value of this charge. Notice that we do not impose any
special constraints to the (real) values of the vacuum charges $\sigma\in \mathbb{R}$.

The Hilbert space $\mc H^\sigma$ is constructed by the action of the negative bosonic generators
\eq{
J_{-n_1}\ldots J_{-n_k}|\sigma\rangle
}
on the vacuum vector $J_0|\sigma\rangle=\sigma|\sigma\rangle$, and these states can be labeled by the Young diagrams with the row lengths $n_1,\ldots,n_k$.

One can also construct the action of the fermionic operators on this vector space. Then the bosonization formulas \rf{bosonization_} will generally produce the fractional powers in holomorphic coordinate $z$ due to the factors $z^{J_0}$, while $e^{\pm Q}$ just shift the vacuum charge by $\pm 1$. It means that one can define the (multiple) action of the modes of the operators
\eq{
\psi^\sigma(z)=\Sum_r\frac{\psi_r^{\sigma}}{z^{r+1/2+\sigma}},\ \ \ \
\tilde{\psi}^\sigma(z)=\Sum_r\frac{\tilde{\psi}^{\sigma}_r}{z^{r+1/2-\sigma}}
}
in the direct sum of the Hilbert spaces
\eq{
\mc H^{\sigma} = \bigoplus_{n\in\mathbb{Z}}\mc H^\sigma_{n}
}
naturally labeled by some fractional $\sigma\in \mathbb{R}/\mathbb{Z}$.

Basis in the each space $\mc H^{\sigma}_n$ can be given by the vectors generated by the zero-charge expressions of the fermionic modes. As in bosonic representation, these vectors can be labeled by the Young diagrams
\eq{
|Y,\sigma\rangle=\prod_i\tilde\psi_{-p_i}^{\sigma}\psi_{-q_i}^{\sigma}|\sigma\rangle
}
where now $p_i$ and $q_i$ are the Frobenius coordinates of the Young diagram. In our convention they are half-integer, and can be easily read of the following picture:\\

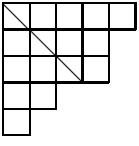
\begin{picture}(60,70)
\put(10,70){\BOXES{0}{-10}{5}\BOXES{0}{-20}{4}\BOXES{0}{-30}{4}\BOXES{0}{-40}{2}\BOXES{0}{-50}{1}
\put(0,0){\line(1,-1){30}}
}
\end{picture}

\noindent
i.e. one has to cut the diagram by the main diagonal and just take the areas of the rows and columns starting from the diagonal cells. For example, the Young diagram from the picture has $\{p_i\}=\{\frac92,\frac52,\frac32\}$ and $\{q_i\}=\{\frac92,\frac52,\frac12\}$.

The states in the dual to $\mc H^{\sigma}$ module can be obtained by the Hermitian conjugation
\eq{
\langle\sigma,Y|= \langle\sigma|\prod_i\tilde\psi_{q_i}^{\sigma}\psi_{p_i}^{\sigma}\,.
}
Our main aim in what follows is to compute the matrix elements of the operator $V_\nu(1)=V_\nu$
between the arbitrary fermionic states
\eq{
Z(\nu|Y',Y)= \langle\theta+\nu,Y'|V_\nu(1)|Y,\theta\rangle\,.
}
The most straightforward way is to use explicit bosonic representation \rf{vertex_} of the vertex operator
\eq{
Z(\nu|Y',Y)=\langle\sigma+\nu|\prod_j\tilde\psi_{q_j'}\psi_{p_j'}V_\nu^-V_\nu^+e^{\nu Q}
\prod_i\tilde\psi_{-p_i}\psi_{-q_i}|\sigma\rangle=
\\
= \langle0|\prod_jV_{-\nu}^-\tilde\psi_{q_j'}V_\nu^-\cdot V_{-\nu}^-\psi_{p_j'}V_\nu^-
\prod_i V_\nu^+\tilde\psi_{-p_i}V_{-\nu}^+\cdot V_\nu^+\psi_{-q_i}V_{-\nu}^+|0\rangle =
\\
= \langle0|\prod_j(V_\nu^-)^{-1}\tilde\psi_{q_j'}V_\nu^-\cdot(V_\nu^-)^{-1}\psi_{p_j'}V_\nu^-
\prod_i V_\nu^+\tilde\psi_{-p_i}(V_\nu^+)^{-1}\cdot V_\nu^+\psi_{-q_i}(V_\nu^+)^{-1}|0\rangle\,.
\label{Zconj}
}
It is easy to understand from \rf{bosonization_} and \rf{vertex_} that the consequent triple products of operators in this formula can be considered as certain adjoint action, or just conjugations of the fermions, which turn under such action just into the linear combinations
of themselves. At the level of generating functions it looks like
\be
\label{adact}
V_\nu^+\tilde{\psi}(z)(V_\nu^+)^{-1}=(1-z)^\nu\tilde{\psi}(z)\,,
\hspace{1cm}
V_\nu^+\psi(z)(V_\nu^+)^{-1}
=(1-z)^{-\nu}\psi(z)\,,\\
(V_\nu^-)^{-1}\tilde{\psi}(z)V_\nu^-=\left(1-\frac1z\right)^{\nu}\tilde{\psi}(z)\,,
\hspace{1cm}
(V_\nu^-)^{-1}\psi(z)V_\nu^-
=\left(1-\frac1z\right)^{-\nu}\psi(z)\,,
\ee
or, more generally
\eq{
V_{\nu}(w)^{-1}\tilde{\psi}^{\sigma+\nu}(z)V_\nu(w)=\left(\frac zw\right)^\nu\exp\left(\nu{\Sum_{n\in\mathbb{Z}}}'\frac1n\frac{z^n}{w^n}\right)
\tilde{\psi}^\sigma(z)\,,
\\
V_{\nu}(w)^{-1}\psi^{\sigma+\nu}(z)V_\nu(w)=\left(\frac zw\right)^{-\nu}\exp\left(-\nu{\Sum_{n\in\mathbb{Z}}}'\frac1n\frac{z^n}{w^n}\right)\psi^\sigma(z)\,,
}
where the formal series in the r.h.s. can be rewritten with the help of the
Fourier transformation as
\be
\exp\left(\nu{\Sum_{n\in\mathbb{Z}}}'\frac{ z^n}{n}\right)=\frac{\sin\pi\nu}{\pi}\Sum_{k\in\mathbb{Z}}\frac{z^k}{k+\nu}\,.
\ee
This is a particular case of transformations from $GL(\infty)$, realized by
\be
\sum a_{rs}:\tilde{\psi}_{-r}\psi_s\ :\in \mathfrak{gl}(\infty),\ \ \ \ a_{rs}\to\infty,\ \  |r-s|\to\infty\,,
\ee
moreover, corresponding to the situation, when $a_{rs}=a_{r-s}$ (a well known example of such transformation
is generated by the currents $J_n = \sum_r :\tilde{\psi}_r\psi_{n-r}:$ from \rf{cur}). It is true in the most general case:
if one computes any matrix elements
of such operator, they always can be expressed in terms of those with only two extra fermion  insertions, i.e. we do not need an explicit form of the operator
$V_\nu=V_\nu^-V_\nu^+$ -- just the only fact of the adjoint action, and we are going to use this property in more complicated non Abelian situation below.

In particular, one can compute \rf{Zconj} first using the Wick theorem
\eq{
Z(\nu|Y',Y)=\det\bpm\langle\sigma+\nu|\tilde\psi_{q_j'}\psi_{p_j'}V_\nu|\sigma\rangle&
\langle\sigma+\nu|\tilde\psi_{q_j'}V_\nu\psi_{-q_i}|\sigma\rangle\\
-\langle\sigma+\nu|\psi_{p_j'}V_\nu\tilde\psi_{-p_i}|\sigma\rangle&\langle\sigma+\nu|
V_\nu\tilde\psi_{-p_i}\psi_{-q_i}|\sigma\rangle\epm=\det G_\nu
\label{det_}
}
and then to apply \rf{adact} to the matrix elements in \rf{det_}.

\subsection{Matrix elements and Nekrasov functions}

The two-fermion matrix elements of the matrix $G=G_\nu$  (its rows are labeled by $\{x_a\}=\{q'_j\}\cup\{-p_i\}$, whereas columns are labeled by  $\{y_b\}=\{p'_j\}\cup\{-q_i\}$, here we denote by $p$ and $q$ some positive half-integer numbers) are expressed as
\be
\label{twopoint}
G(q',p')=\langle0|\psi_{q'}\tilde{\psi}_{p'}V_\nu^-|0\rangle=
\Sum_{m=0}^{q'-\frac12}\frac{(\nu)_m}{m!}\frac{(-\nu)_{p'+q'-m}}{(p'+q'-m)!}\,,
\\
G(-p,-q)=\langle0|V_\nu^+\psi_{-p}\tilde{\psi}_{-q}|0\rangle=
\Sum_{n=0}^{q-\frac12}\frac{(-\nu)_n}{n!}\frac{(\nu)_{p+q-n}}{(p+q-n)!}\,,
\\
G(-p,p')=-\langle0|\tilde{\psi}_{p'}V_\nu^-V_\nu^+\psi_{-p}|0\rangle=-\Sum_{m=0}^{p'-\frac12}
\frac{(-\nu)_m}{m!}\frac{(\nu)_{m+p-p'}}{
(m+p-p')!}\,,\\
G(q',-q)=\langle0|\psi_{q'}V_\nu^-V_\nu^+\tilde{\psi}_{-q}|0\rangle=
\Sum_{n=0}^{q-\frac12}\frac{(-\nu)_n}{n!}\frac{(\nu)_{n+q'-q}}{(n+q'-q)!}\,.
\ee
These expressions are easily computed, using adjoint action \rf{adact} for the components
\be
V_\nu^+\psi_{-p}(V_\nu^+)^{-1}=\Sum_{m=0}^\infty\frac{(\nu)_m}{m!}\psi_{-p+m},\ \ \ \
V_\nu^+\tilde{\psi}_{-q}(V_\nu^+)^{-1}=\Sum_{m=0}^\infty\frac{(-\nu)_m}{m!}\tilde{\psi}_{-q+m}\\
(V_\nu^-)^{-1}\psi_{q}V_\nu^-=\Sum_{m=0}^\infty\frac{(\nu)_m}{m!}\psi_{q-m},\ \ \ \
(V_\nu^-)^{-1}\tilde{\psi}_{p}V_\nu^-=\Sum_{m=0}^\infty\frac{(-\nu)_m}{m!}\tilde{\psi}_{p-m}
\ee
with $(\nu)_m = \nu(\nu+1)\ldots(\nu+m-1)$, $(\nu)_0=1$, and there are explicit formulas for the sums in the r.h.s. of \rf{twopoint}
\be
\Sum_{m=0}^b\frac{(\nu)_m}{m!}\frac{(-\nu)_{a-m}}{(a-m)!}=\frac{(\nu)_{b+1}(-\nu)_{a-b}}{\nu a b!(a-b-1)!}
\\
\Sum_{m=0}^b\frac{(-\nu)_m}{m!}\frac{(\nu)_{a+m}}{(a+m)!}=-\frac{(-\nu)_{b+1}(\nu)_{a+b+1}}{\nu(a+\nu)b!(a+b)!}
\ee
which can be easily proven by induction. It allows to rewrite matrix elements \rf{twopoint}
in the factorized form
\be
G(q',p')=\frac1{\nu(p'+q')}\frac{(\nu)_{q'+\frac12}(-\nu)_{p'+\frac12}}{(q'-\frac12)!(p'-\frac12)!}\,,\\
G(-p,-q)=-\frac1{\nu(p+q)}\frac{(\nu)_{p+\frac12}(-\nu)_{q+\frac12}}{(p-\frac12)!(q-\frac12)!}\,,\\
G(-p,p')=\frac1{\nu(p-p'+\nu)}\frac{(\nu)_{p+\frac12}(-\nu)_{p'+\frac12}}{(p-\frac12)!(q'-\frac12)!}\,,\\
G(q',-q)=-\frac1{\nu(q'-q+\nu)}\frac{(\nu)_{q'+\frac12}(-\nu)_{q+\frac12}}{(q-\frac12)!(q'-\frac12)!}\,.
\label{mef}
\ee
The determinant from \rf{det_} can be therefore written as
\be
\label{wick2}
\det_{a,b} G(x_a,y_b)=
\prod_{j}\frac{(-\nu)_{p'_j+\frac12}(\nu)_{q'_j+\frac12}}{\nu(p'_j-\frac12)!(q'_j-\frac12)!}
\prod_i\frac{(\nu)_{p_i+\frac12}(-\nu)_{q_i+\frac12}}{\nu(p_i-\frac12)!(q_i-\frac12)!}\cdot
\det_{a,b} \tilde{G}(\tilde{x}_a,\tilde{y}_b)
\ee
where now for two new sets $\{\tilde{x}_a\}=\{q_j'\}\cup\{-p_i-\nu\}$, $\{\tilde{y}_b\}=\{-p_j'\}\cup\{q_i-\nu\}$
\be
\tilde{G}(\tilde{x}_a,\tilde{y}_b) = {{\rm sgn}(\tilde{x}_a\tilde{y}_b)\over \tilde{x}_a-\tilde{y}_b}\,,
\ee
and the corresponding determinant can be computed using the Cauchy determinant formula
$$\det_{a,b}\frac1{\tilde{x}_a-\tilde{y}_b}=
\frac{\prod_{a<b}(\tilde{x}_a-\tilde{x}_b)\prod_{a>b}(\tilde{y}_a-\tilde{y}_b)}
{\prod_{ab}(\tilde{x}_a-\tilde{y}_b)}\,,$$
so one gets finally
\be
\label{mael}
Z(\nu|Y',Y)=\pm\prod_{j}
\frac{(-\nu)_{p'_j+\frac12}(\nu)_{q'_j+\frac12}}{\nu(p'_j-\frac12)!(q'_j-\frac12)!}
\prod_i\frac{(\nu)_{p_i+\frac12}(-\nu)_{q_i+\frac12}}{\nu(p_i-\frac12)!(q_i-\frac12)!}\times\\\times
\frac{\prod_{i>j}(p_i'-p_j')\prod_{i<j}(p_i-p_j)\prod_{i>j}(q'_i-q'_j)\prod_{i<j}(q_i-q_j)\prod_{ij}
(q_i'+p_j+\nu)\prod_{ij}(p_i'+q_j-\nu)}{\prod_{ij}(p'_i+q'_j)\prod_{ij}(p_i+q_j)
\prod_{ij}(q'_i-q_j+\nu)\prod_{ij}(p_i-p_j'+\nu)}
\ee
It is easy to see that this expression has the structure
\be
Z(\nu|Y',Y)=\pm\frac{Z_b(\nu|Y',Y)}{Z_0^\frac12(Y')Z_0^\frac12(Y)}
\ee
where
\be
Z_0^\frac12(Y)=\prod_i\left(p_i-\half\right)!\left(q_i-\half\right)!\frac{\prod_{ij}(p_i+q_j)}
{\prod_{i<j}(q_i-q_j)\prod_{i<j}(p_i-p_j)}\,,
\ee
while
\be
Z_b(\nu|Y',Y)=\prod_i\nu^{-1}(-\nu)_{p_i'+\frac12}(\nu)_{q_i'+\frac12}
\prod_j\nu^{-1}(-\nu)_{q_j+\frac12}(\nu)_{p_j+\frac12}
\times
\\
\times\frac{\prod_{ij}(q_i'+p_j+\nu)\prod_{ij}(p_i'+q_j-\nu)}
{\prod_{ij}(q_i'-q_j+\nu)\prod_{ij}(p_i'-p_j-\nu)}\,.
\label{Zb}
\ee
In this normalization one can check that
\be
Z_b(\nu|Y',Y)=\pm\prod_{t\in Y}(1+a_{Y}(t)+l_{Y'}(t)+\nu)\prod_{s\in Y'}(1+a_{Y'}(s)+l_{Y}(s)-\nu)
\label{ZbN}
\ee
is exactly the Nekrasov bi-fundamental function of the $U(1)$ gauge theory at $c=1$ or $\epsilon_1+\epsilon_2=0$.
Notice also that
\be
Z_b(0|Y,Y)=Z_0^\frac12(Y)Z_0^\frac12(Y) = \prod_{s\in Y}(1+a_{Y}(s)+l_{Y}(s))^2 = Z_V(Y)^{-1}
\ee
is Nekrasov function for the pure $U(1)$ gauge theory, which corresponds to the Plancherel
measure on partitions  \cite{LMN}.

\subsection{Riemann-Hilbert problem}

The following simple observation is extremely important for our generalizations below.
Consider the correlator
\be
\langle\theta|V_\nu(1)\tilde\psi^\sigma(z)\psi^\sigma(w)|\sigma\rangle=
\delta_{\theta,\sigma+\nu}\frac{z^\sigma w^{-\sigma}(1-z)^\nu (1-w)^{-\nu}}{z-w}
\label{ppcorr}
\ee
which is easily computed using bosonization rules \rf{bosonization_}. One finds then, that
\be
(z-w)\langle\theta|V_\nu(1)\tilde\psi^\sigma(z)\psi^\sigma(w)|\sigma\rangle=\phi(z)\phi(w)^{-1}
\ee
is expressed actually through the solutions of a simple linear system
\eq{
\frac{d\phi(z)}{dz}=\phi(z)\left(\frac{\sigma}{z}+\frac{\nu}{z-1}\right)
}
It means that this linear system can be used to define all two-fermion matrix elements, e.g. in the region $1>|z|>|w|$
\eq{
\langle\theta|V_\nu(1)\tilde\psi^\sigma(z)\psi^\sigma(w)|\sigma\rangle=\Sum_{p,q}\frac1{z^{p+\frac12-\sigma}w^{q+\frac12+\sigma}}
\langle\theta|V_\nu(1)\tilde\psi^\sigma_p\psi^\sigma_q|\sigma\rangle
}
and together with the Wick theorem it defines all matrix elements, or just the vertex operator $V_\nu$, uniquely -- up to a numeric factor. In its turn the linear system itself is determined by the monodromy properties
(here very simple) of $\phi(z)$ at $z=0$ and $z=1$ (and related to them monodromy at $z=\infty$). Hence, the problem of computation of the two-fermion correlation functions can be reformulated in terms of a Riemann-Hilbert problem.

\subsection{Remarks}

\begin{itemize}
  \item Formulas \rf{mael}, \rf{Zb} give a very explicit representation for the matrix element and bi-fundamental Nekrasov function in terms of the Frobenius coordinates of the corresponding Young diagrams (this
representation, for example, is far more adapted for practical computation, than the formulas \rf{ZbN}).
However, it is sometimes not easy to see directly, that these formulas possess some nice properties:
for example satisfy the ``sum rules'' like
\eq{
\Sum_Y
t^{|Y|}\frac{Z_b(\alpha_1|\emptyset,Y)Z_b(\alpha_2|Y,\emptyset)}{Z_0(Y)}=
\Sum_Y t^{|Y|}Z(\alpha_1|\emptyset,Y)Z(\alpha_2|Y,\emptyset) =
\\
=\Sum_Y t^{|Y|}\langle0|e^{i\alpha_1\phi(1)}|Y,0\rangle\langle Y,0|e^{i\alpha_2\phi(1)}|0\rangle = (1-t)^{\alpha_1\alpha_2}
}
where the r.h.s. immediately follows from resolution of unity and
the correlator of two exponentials
\eq{
\langle 0|e^{i\alpha_1\phi(1)}e^{i\alpha_2\phi(t)}|0\rangle=(1-t)^{\alpha_1\alpha_2}
}
which is instructive to compare with the computation from \cite{NSlav,Kozlowski}.
\item One can also easily extract some useful information from particular cases of \rf{ppcorr},
    which include a nice identity (cf. with \cite{NO,Wiegmann})
\eq{
\frac{(1-z)^{\nu}(1-w)^{-\nu}}{z-w}=\frac1{z-w}+\Sum_{a,b=0}^\infty\frac{(-\nu)_{a+1}(\nu)_{b+1}}{(a+b+1)\nu a!b!}z^aw^b
\label{identity1_}
}
containing some part of the matrix elements from \rf{mef}.
  \item According to \rf{no}
\be
\tilde{\psi}(z+t/2)\psi(z-t/2) = {1\over t}:\exp\left(\int_{z-t/2}^{z+t/2} J(\xi)d\xi\right):
\ee
Expansion into the powers of $t$ gives the infinite series of the currents of $W_{1+\infty}$ algebra
\be
:\tilde{\psi}(z+t/2)\psi(z-t/2): = {1\over t}:\left(\exp\left(\int_{z-t/2}^{z+t/2} J(\xi)d\xi\right)-1\right):
=
\\
= \sum_{k>0}{t^{k-1}\over (k-1)!}U_k(z)
\label{POP}
\ee
where explicitly
\be
U_1(z) = J(z),\ \ \ U_2= \half:J(z)^2:,\ \ \ U_3 ={1\over 3}\left(:J(z)^3: + {1\over 4}\pd^2J(z)\right),\ \ \ \ldots
\ee
and one implies bosonic normal ordering for the bosons and fermionic for the fermions. These formulas
have been used many times (see e.g. \cite{Pogr,OP,LMN,NO,PoPya}) to relate the generators of the $W_{1+\infty}$ algebra with the fermionic bilinear operators, and we just recall them in order to generalize below to much less trivial non Abelian case.

\end{itemize}

\setcounter{equation}0
\section{Non-Abelian $U(N)$ theory}

\subsection{Nekrasov functions}

Consider now more general case of Nekrasov functions, corresponding to the $U(N)$ non-Abelian theory. They can be expressed in terms of $U(1)$ functions \rf{Zb}, \rf{ZbN} by the following product formula
\be
\hat Z_b(\bs\theta',\nu,\bs\theta|\bs Y',\bs Y)=
\Prod_{\alpha,\beta=1}^NZ_b(\nu-\theta'_\alpha+\theta_\beta|Y'_\alpha,Y_\beta)
\label{ZbUN}
\ee
For the diagonal elements $\hat Z_0(\bs\theta|\bs Y)=\hat Z_b(\bs\theta,0,\bs\theta|\bs Y,\bs Y)$ one gets
\be
\hat Z_0(\bs\theta|\bs Y)=
\Prod_{\alpha,\beta=1}^NZ_b(-\theta_\alpha+\theta_\beta|Y_\alpha,Y_\beta)=\pm\prod_{i<j}Z^2_b(-\theta_\alpha+
\theta_\beta|Y_\alpha,Y_\beta)\cdot\Prod_\alpha Z_0(Y_\alpha)
\ee
or, after taking the square root, just
\eq{
\hat Z^\frac12_0(\bs\theta|\bs Y)=\prod_{\alpha<\beta}Z_b(-\theta_\alpha+\theta_\beta|Y_\alpha,Y_\beta)
\cdot\Prod_\alpha Z^\frac12_0(Y_\alpha)
}
Now for simplicity it is better to replace $\theta'_\alpha-\nu\mapsto \theta'_\alpha$ or $\theta_\alpha+\nu\mapsto\theta_\alpha$, then $\nu$ simply disappears from \rf{ZbUN}.
Consider now the normalized matrix element
\be
\hat Z(\bs\theta',\bs\theta|\bs Y',\bs Y)=
\frac{\hat Z_b(\bs\theta',\bs\theta|\bs Y',\bs Y)}{\hat Z_0^\frac12(\bs\theta'|\bs Y')\hat Z_0^\frac12(\bs\theta|\bs Y)}=\\
=\frac{\Prod_{\alpha,\beta=1}^NZ_b(-\theta'_\alpha+\theta_\beta|Y'_\alpha,Y_\beta)}{
\prod_{\alpha<\beta}Z_b(-\theta_\alpha+\theta_\beta|Y_\alpha,Y_\beta)\Prod_\alpha Z^\frac12_0(Y_\alpha)\cdot
\prod_{\alpha<\beta}Z_b(-\theta'_\alpha+\theta'_\beta|Y'_\alpha,Y'_\beta)
\cdot\Prod_\alpha Z^\frac12_0(Y'_\alpha)}
\label{UNfromU1}
\ee
Using representation \rf{mael}, \rf{Zb} for the $U(1)$ functions in terms of the Frobenius coordinates,
one finds that the ratio of products of the elementary Cauchy determinants from there is actually combined into more sophisticated unique Cauchy determinant
\be
\hat Z(\bs\theta',\bs\theta|\bs Y',\bs Y)= \det_{IJ}\frac1{x_I-y_J}\times
\\
\times\Prod_{i,\alpha} f_{1,\alpha}(\bs\theta',\bs\theta,p'_{\alpha,i})
f_{2,\alpha}(\bs\theta',\bs\theta,q'_{\alpha,i}) f_{1,\alpha}(\bs\theta,\bs\theta',p_{\alpha,i})
f_{2,\alpha}(\bs\theta,\bs\theta',q_{\alpha,i})
\label{UNdet}
\ee
with two multi-sets of variables entering the determinant of the form
\be
\{x_I\}=\{-q'_{\alpha, i}-\theta_{\alpha}'\}\cup\{p_{\alpha,i}-\theta_\alpha\}\\
\{y_I\}=\{p_{\alpha,i}'-\theta'_\alpha\}\cup\{-q_{\alpha,i}-\theta_\alpha\}
\ee
up to quite nontrivial diagonal part, which can be still read from \rf{mael} and
\rf{UNfromU1}, giving the following factors for \rf{UNdet}
\be
f_{1,\alpha}(\bs\theta,\bs\theta',p_{\alpha,i})=
\frac1{(p_{\alpha,i}-\frac12)!}
\Prod_\beta\frac{(\theta'_\beta-\theta_\alpha)_{p_{\alpha,i}+\frac12}}{\sqrt{\theta'_\beta-\theta_\alpha}}
\prod_{\beta\neq\alpha}\frac{\sqrt{\theta_\beta-\theta_\alpha}}{(\theta_\beta-\theta_\alpha)_{p_{\alpha,i}+\frac12}}
\\
f_{2,\alpha}(\bs\theta,\bs\theta',q_{\alpha,i})=
\frac1{(q_{\alpha,i}-\frac12)!}
\Prod_\beta\frac{(\theta_\alpha-\theta'_\beta)_{q_{\alpha,i}+\frac12}}{\sqrt{\theta_\alpha-\theta'_\beta}}
\prod_{\beta\neq\alpha}\frac{\sqrt{\theta_\alpha-\theta_\beta}}{(\theta_\alpha-\theta_\beta)_{q_{\alpha,i}
+\frac12}}
\label{fprod}
\ee
Existence of the determinant formula \rf{UNdet} is very important, since it actually implies that Nekrasov functions $\hat Z(\bs\theta',\bs\theta|\bs Y',\bs Y)$
can be identified with the matrix elements of some vertex operator, characterized as in the Abelian $U(1)$ case by its adjoint action, which is still a linear transformation but now of the $N$-component fermions.
We are going indeed to introduce this vertex operator below using the theory of ($N$-component) free fermions, generalizing the Abelian case considered above.
In general situation this operator is characterized by solution to auxiliary linear problem on sphere
with three marked points, while explicit formulas
of this section just correspond to particular case of the hypergeometric-type solutions.

\subsection{$N$-component free fermions}

Hence, consider the generalization of the free-fermionic construction from $U(1)$ to the non-Abelian $U(N)$ case. First, introduce the algebra
\be
\{\psi_{\alpha,r},\psi_{\beta,s}\}=0\,,\qquad \{{\tilde\psi}_{\alpha,r},
{\tilde\psi}_{\beta,s}\}=0\,,\\
\{{\tilde\psi}_{\alpha,r},\psi_{\beta,s}\}=\delta_{\alpha,\beta}\delta_{r+s,0}
\\
r,s\in \mathbb{Z}+\half,\ \ \ \ \alpha,\beta=1,\ldots,N
\label{malgs}
\ee
of the canonical anticommutation relations for the components of the fermionic fields
with free first-order action $S = {1\over\pi}\sum_{\alpha=1}^N\int_\Sigma d^2z \tilde{\psi}_\alpha\d\psi_\alpha$, so that \rf{malgs} are equivalent to the operator product expansions
\be
{\tilde{\psi}}_\alpha(z)\psi_\beta(w)=\frac{\delta_{\alpha\beta}}{z-w}+J_{\alpha\beta}(w) + O(z-w)
\\
\psi_\alpha(z)\psi_\beta(w)=reg.\qquad
{\tilde{\psi}}_\alpha(z){\tilde{\psi}}_\beta(w)=reg.
\label{NOPE}
\ee
Similarly to \rf{bosonization_}
it is also possible and useful to introduce the bosonization formulas for these fermionic fields
\eqs{
\tilde\psi_\alpha(z)=&\exp\left(-\Sum_{n<0}\frac{J_{\alpha,n}}{nz^{n}}\right)\exp
\left(-\Sum_{n>0}\frac{J_{\alpha,n}}{nz^{n}}\right)e^{Q_\alpha} z^{J_{\alpha,0}}\epsilon_\alpha(\bs J_0)\\
\psi_\alpha(z)=&\exp\left(\Sum_{n<0}\frac{J_{\alpha,n}}{nz^{n}}\right)\exp
\left(\Sum_{n>0}\frac{J_{\alpha,n}}{nz^{n}}\right)e^{-Q_\alpha} z^{-J_{\alpha,0}}\epsilon_\alpha(\bs J_0)\label{bosonization}
}
Here $J_{\alpha,n}$ form the Heisenberg algebra
\eq{
\left[J_{\alpha,n},J_{\beta,m}\right]=n\delta_{\alpha\beta}\delta_{m+n,0},
\ \ \ \ \ \ \ [J_{0,\alpha},Q_\beta]=\delta_{\alpha\beta}
\label{heisenberg}
}
and $\epsilon_\alpha(\bs J_0)=\prod_{\beta=1}^{\alpha-1}(-1)^{J_{0,\beta}}$, we may also note that $\epsilon_\alpha(\bs x+\bs y)=\epsilon_\alpha(\bs x)\epsilon_\beta(\bs y)$. These extra sign factors do the same as the Jordan-Wigner transformation:
they convert commuting objects into the anticommuting ones.

A standard representation of this algebra $\mc H^{\bs\sigma}$ is constructed from the vacuum vector $|\bs\sigma\rangle$, with the charges ${\bs J_0}|\bs\sigma\rangle = {\bs \sigma}|\bs\sigma\rangle$ and killed by all positive modes
\be
\psi^{\bs\sigma}_{\alpha,r>0}|\bs\sigma\rangle=0\,,\qquad {\tilde\psi}^{\bs\sigma}_{\alpha,r>0}|\bs\sigma\rangle=0\,.
\ee
Basis vectors of this representation can be given by
\eq{|\{p_{\alpha,i}\},\{q_{\alpha,i}\},\bs\sigma\rangle=\prod_{\alpha=1}^N\left(\prod_{i=1}^{\left| p_{\alpha,i}\right|}\tilde\psi_{\alpha,-p_{\alpha,i}}
\prod_{j=1}^{\left| q_{\alpha,j}\right|}\psi_{\alpha,-q_{\alpha,j}}\right)|\bs\sigma\rangle
\label{sigstates}
}
The letters $p_{\alpha,i}$ and $q_{\alpha,i}$, at least in the case when $\# p_\alpha=\#q_\alpha$, should be interpreted as Frobenius coordinates of
the $N$-tuple of the Young diagrams.
It will be also convenient in what follows to use the vacuum-shifting operators $P^n_\alpha$
\eq{
P^0_\alpha=1,\qquad P^{n<0}_\alpha=\psi^{\bs\sigma}_{\alpha,n+\frac12}\psi^{\bs\sigma}_{\alpha,n+\frac32}
\ldots\psi^{\bs\sigma}_{\alpha,-\frac12}|\bs\sigma\rangle\,,\\
P^{n>0}_\alpha={\tilde\psi}^{\bs\sigma}_{\alpha,-n+\frac12}{\tilde\psi}^{\bs\sigma}_{\alpha,-n+\frac32}
\ldots{\tilde{\psi}}^{\bs\sigma}_{\alpha,-\frac12}|\bs\sigma\rangle\label{shifts}
}
and the corresponding states
\be
|\bs n,\bs\sigma\rangle=\Prod_{\alpha=1}^N P_\alpha^{n_\alpha}|\bs\sigma\rangle\,.
\label{nsig}
\ee
in particular for the vectors $\mathbf{n} = \pm \mathbf{1}_\beta$ with components $n_\alpha=\pm\delta_{\alpha\beta}$.

\subsection{Level one Kac-Moody and W-algebras}

Consider the W-algebras for $\mathfrak{g}=\mathfrak{sl}(N)$ series, possibly extended to
$\mathfrak{gl}(N)$ where we shall call it $W_N\oplus H$.
Their generators in current representation can be identified with the symmetric functions
of the normally ordered currents $J(z)\in \mathfrak{h}\subset \mathfrak{g}$ with the values
in Cartan subalgebra, or equivalently, up to a coefficient, as certain ``Casimir elements''
in the universal enveloping $U(\widehat{\mathfrak{sl}(N)}_1)$. The Virasoro central charge at level $k=1$ is
\be
c = {k\dim \mathfrak{g}\over k+C_V} = {N^2-1\over 1+N}=N-1
\ee
When embedded to $U(\widehat{\mathfrak{gl}(N)}_1)$ this current algebra has nice representation in terms of the
multi-component free holomorphic fermionic fields
\be
J_{\alpha\beta}(z) = :\tilde\psi_\alpha(z)\psi_\beta(z):,\ \ \ \ \alpha,\beta=1,\ldots,N
\ee
The $W_N$-algebra can be defined in terms of invariant Casimir polynomials of the
currents, commuting with the screening charges $Q_{\alpha\beta} = \oint  J_{\alpha\beta}(z)$
(it is enough to require commutativity only with those, corresponding to the positive simple roots).
Then the W-generators turn to be just the symmetric polynomials of the diagonal Cartan currents
$J_\alpha = J_{\alpha\alpha}(z) \in \mathfrak{h}$, i.e.
\be
W_n(z) = \sum_{\alpha_1<\alpha_2<\ldots<\alpha_n} :J_{\alpha_1}(z)J_{\alpha_2}(z)\ldots J_{\alpha_n}(z):\,,
\ \ \ \ \ n=1,\ldots,N
\ee
One can consider the representations of $U(\widehat{\mathfrak{gl}(N)}_1)$ and $W_N\oplus H$ in $\mc H^{\bs\sigma}$.
For this purpose it is convenient to introduce the generating functions
\be
\psi^{\bs\sigma}_\alpha(z)=\Sum_{r\in\mathbb{Z}+\frac12}
\frac{\psi_{\alpha,r}^{\bs\sigma}}{z^{r+\frac12+\sigma_\alpha}}\,,\ \ \ \ \
{\tilde\psi}^{\bs\sigma}_\alpha(z)=\Sum_{r\in\mathbb{Z}+\frac12}
\frac{{\tilde\psi}_{\alpha,r}^{\bs\sigma}}{z^{r+\frac12-\sigma_\alpha}}\,.
\ee
where shifts of the powers of the coordinate $z$ come naturally, e.g. from the bosonization formulas \rf{bosonization}. For these fields instead of \rf{NOPE} one gets
\be
\tilde{\psi}^{\bs\sigma}_\alpha(z)\psi^{\bs\sigma}_\beta(w)=
\delta_{\alpha\beta}\frac{z^{\sigma_\alpha}w^{-\sigma_\alpha}}{z-w}+
:\tilde{\psi}^{\bs\sigma}_\alpha(z)\psi^{\bs\sigma}_\beta(w):\,=
\\
=\frac{\delta_{\alpha\beta}}{z-w} + \delta_{\alpha\beta}\frac{\sigma_\alpha}{w}+
:\tilde{\psi}^{\bs\sigma}_\alpha(w)\psi^{\bs\sigma}_\beta(w):+\, O(z-w)\,,
\ee
then it is clear that the modes of $J_{\alpha\beta}^{\bs\sigma}(w)$ from \rf{NOPE} acquire in
this representation the form
\be
J^{\bs\sigma}_{\alpha\beta,n}=\delta_{\alpha\beta}\delta_{n,0}\sigma_\alpha+
\Sum_{p\in\mathbb{Z}+\frac12}:\tilde{\psi}^{\bs\sigma}_{\alpha,n-p}
\psi^{\bs\sigma}_{\beta,p}:
\ee
As in the $U(1)$ case (see \rf{POP}) in the fermionic realization of $W_N\oplus H$, then one can
choose the set of generators in a form of the fermionic bilinears:
\be
\Sum_\alpha{\tilde\psi}^{\bs\sigma}_\alpha(z+\frac t2)\psi^{\bs\sigma}_\alpha(z-\frac t2)=\frac Nt+\Sum_{k=1}^\infty\frac{t^{k-1}}{(k-1)!}
U_k^{\bs\sigma}(z)\,.\label{Wdef}
\ee
The l.h.s. of this formula gives
\be
\tilde{\psi}^{\bs\sigma}_\alpha(z+\frac t2)\psi^{\bs\sigma}_\alpha(z-\frac t2)=\frac1t\left(\frac{1+\frac t{2z}}{1-\frac t{2z}}\right)^{\sigma_\alpha}+
\\+\frac1t\Sum_{m\in\mathbb{Z}}\frac1{z^m}\frac{\frac{t}{z}}{(1+\frac t{2z})^{m+1}}\Sum_{p\in\mathbb{Z}+\frac12}
\left(\frac{1+\frac t{2z}}{1-\frac t{2z}}\right)^{p+\frac12+\sigma_\alpha}:\tilde{\psi}^{\bs\sigma}_{m-p,\alpha}\psi^{\bs\sigma}_{\alpha,p}:\,.
\ee
Introducing two collections of polynomials~\footnote{One can also notice at the level of the generating functions \rf{genfunc} that
$$v_{k,0}(p)=u_{k}(p+\frac12)-u_{k}(p-\frac12)\,.$$
First polynomials are given explicitly by
\be
u_1(p)=p,\quad
u_2(p)=\frac{p^2}2,\quad
u_3(p)=\frac{p^3}3+\frac p6,\\
u_4(p)=\frac{p^4}4+\frac{p^2}2,\quad
u_5(p)=\frac{p^5}5+p^3+\frac{3p}{10},\quad
\ldots
\ee}
\be
\left(\frac{1+\frac x2}{1-\frac x2}\right)^p=1+\Sum_{k=0}^\infty u_k(p)\frac{x^k}{(k-1)!}\,,\\
\frac{x}{(1+\frac x2)^{m+1}}\left(\frac{1+\frac x2}{1-\frac x2}\right)^{p+\frac12}=\Sum_{k=1}^\infty v_{k,m}(p)\frac{x^k}{(k-1)!}\,,\label{genfunc}
\ee
the generators in the r.h.s. of \rf{Wdef} explicitly become
\be
U^{\bs\sigma}_{k,m}=\Sum_\alpha\left(\delta_{m,0} u_k(\sigma_\alpha)+\Sum_{p\in\mathbb{Z}}
v_{k,m}(p+\sigma_\alpha):\tilde{\psi}^{\bs\sigma}_{\alpha,m-p}\psi^{\bs\sigma}_{p,\alpha}:\right).
\ee
This set of generators of $W_N\oplus H$ contains commuting zero modes $U^{\bs\sigma}_{k,0}$ which were
shown to play an important role in the study of the extended Seiberg-Witten theory and AGT correspondence \cite{LMN,MN,PoPya,Litv3}.
It is also important to notice
that commutation relations between these generators are linear, the only place when the
non-linearity appears are the relations between these generators.

Using the bosonization rules \rf{bosonization} one can rewrite these generators in the conventional form.
To perform explicit splitting of this algebra into $W_N\oplus H$ it is convenient to redefine
$J_\alpha(z)\mapsto J_\alpha(z)+j(z)$, where the new currents already satisfy the condition $\Sum J_\alpha=0$ and the operator product expansions (OPE)
\be
j(z)j(w)=\frac{\frac1N}{(z-w)^2}+reg.\qquad
J_\alpha(z)J_\beta(w)=\frac{\delta_{\alpha\beta}-\frac1N}{(z-w)^2}+reg.
\ee
Now we take the bilinear expression
\be
\Sum_\alpha\tilde\psi_\alpha(z+\frac t2)\psi_\alpha(z-\frac t2)=\Sum_\alpha\,:e^{i\varphi(z+\frac t2)+i\phi_\alpha(z+\frac t2)}:
:e^{-i\varphi(z-\frac t2)-i\phi_\alpha(z-\frac t2)}:\,=\\=\frac1t:e^{i\varphi(z+\frac t2)-i\varphi(z-\frac t2)}:
\Sum_\alpha:e^{i\phi_\alpha(z+\frac t2)-i\phi_\alpha(z-\frac t2)}:
\label{psipsi}
\ee
with $j(z)=i\pd\varphi$, $J_\alpha(z)=i\pd \phi_\alpha(z)$ and expand it into the powers of $t$.
Comparing with \rf{Wdef} we get the following formulas:
\be
U_1(z)=N j(z),\qquad U_2(z)=T(z)+\frac N2 :j^2(z):,\\
U_3(z)=W_3(z)+2NT(z)j(z)+\frac N3\left(:j^3(z):+\frac14\pd^2 j(z)\right),\\
U_4(z)=-W_4(z)+\frac12(TT)(z)+3 W_3(z)j(z)+3 :j^2(z):T(z)+\\+\frac N4\left(:j^4(z):+:j(z)\pd^2j(z):\right),
\qquad U_5(z)=\ldots
\label{UjT}
\ee
where $T(z)=-W_2(z)$ is the stress-energy tensor, and $(AB)(z)$ is the ``interacting'' normal ordering
$$
(AB)(z)=\oint_z\frac{dw}{w-z}A(w)B(z)
$$
One find therefore, that one basis is related with the other by some complicated, though explicit and triangular transformation. Here we can see that generators $U_k(z)$ are actually dependent, namely, if $N=3$, then
$W_4(z)=0$ and $U_4(z)$ becomes some non-linear expression of the lower generators.

It is also easy to see that for the states \rf{nsig}
\be
J^{\bs\sigma}_{\alpha,0}|\bs n,\bs\sigma\rangle=(\sigma_\alpha+n_\alpha)|\bs n,\bs\sigma\rangle\,,\ \ \ \
U^{\bs\sigma}_{k,0}|\bs n,\bs\sigma\rangle=u_k(\bs\sigma+\bs n)|\bs n,\bs\sigma\rangle\,,\\
U^{\bs\sigma}_{k,m>0}|\bs n,\bs\sigma\rangle=0.
\label{highestweight}
\ee
It is sometimes useful to decompose the whole Hilbert space into the sectors
$\mc H^{\bs\sigma}=\bigoplus\limits_{\bs n\in\mathbb{Z}^N}\mc H^{\bs\sigma}_{\bs n}$
with fixed $\mathfrak{h}\in\mathfrak{gl}(N)$
charges and also into the  sectors
$\mc H^{\bs\sigma}_l=\bigoplus\limits_{\Sum n_\alpha=l}\mc H^{\bs\sigma}_{\bs n}$
with fixed overall $u(1)=\mathfrak{gl}(1)$ charge.
Summarizing all these facts we can formulate the following
\begin{theorem}\label{thm:highest_weight}
Spaces $\mc H_l^{\bs\sigma}$ are representations of $\widehat{\mathfrak{gl}(N)}_1$, and  for
general $\bs\sigma$ spaces $\mc H_{\bs n}^{\bs\sigma}$ are the Verma modules of $W_N\oplus H$ algebra with the highest weight vectors $|\bs\sigma,\bs n\rangle$ and
with basis vectors $|\bs Y,\bs n, \bs\sigma\rangle$, $\forall{\bs Y}$.
\end{theorem}
{\bf Proof} is extremely simple: $\widehat{\mathfrak{gl}(N)}_1$  generators have zero fermionic $\mathfrak{gl}_1$-charge, $W_N\oplus H$ generators have
zero  charges with respect to the whole Cartan subalgebra $\mathfrak{h}$, so the spaces  $\mc H_l^{\bs\sigma}$ and $\mc H_{\bs n}^{\bs\sigma}$ are closed under the action
of these algebras. We also know from \rf{highestweight} that $|\bs\sigma,\bs n\rangle$ are the highest weight vectors
of $W_N\oplus H$, so we have a non-zero map from the Verma module to $\mc H_{\bs n}^{\bs\sigma}$, but this Verma module is generally
irreducible and has the same character $\tr q^{L_0}$, so we actually have an isomorphism. \hfill $\square$

\subsection{Free fermions and representations of W-algebras}

Let us now illustrate how can free fermions appear in the theory with $W_N$-symmetry at integer central charges after inclusion of extra Heisenberg algebra. Construction below is a straightforward generalization of the bosonization procedure from \cite{ILT}.

It is well-known \cite{FZ,FL} that conformal theory with $W_N$-symmetry contains two degenerate fields  $V_{\mu_1}(z)$ and $V_{\mu_{N-1}}(z)$, such that their
$W$-charges are determined by the highest weights of the fundamental ($\mathbf{N}$) and antifundamental ($\bar{\mathbf{N}}$) representations, respectively.  Their dimensions are
\eq{\Delta(\mu_i)=\half \mu_i^2 = \frac{N-1}{2N},\ \ \ \ \ i=1,N-1}
and they have the following fusion rules with arbitrary primary field
\eqs{\left[\mu_1\right]\otimes[\bs\sigma]&=\oplus_{\alpha=1}^N[\bs\sigma+e_\alpha]\\
[\mu_{N-1}]\otimes[\bs\sigma]&=\oplus_{\alpha=1}^N[\bs\sigma-e_\alpha]
\label{mufus}}
where $\{\pm e_\beta\}$ is the set of all weights of $\mathbf{N}$ and $\bar{\mathbf{N}}$.
One can define now the vertex operators
\eq{\Psi_\alpha(z)=\sum_{\bs\sigma}P_{\bs\sigma+ e_\alpha}V_{\mu_1}(z)P_{\bs\sigma},\ \ \ \
\tilde\Psi_\alpha(z)=\sum_{\bs\sigma}P_{\bs\sigma-e_\alpha}V_{\mu_{N-1}}(z)P_{\bs\sigma}
}
which, due to extra projector operators, act only from one Verma module to another, just extracting the corresponding term from the fusion rules \rf{mufus}. Using the general structure of the
OPE of two initial degenerate fields
\be
V_{\mu_1}(z)V_{\mu_{N-1}}(w)=\left(\mathbf{1}\cdot(z-w)^{\frac{1-N}{N}}+\# \cdot(z-w)^{\frac{1+N}{N}}T(w)\right)+
\\
+ (z-w)^{\frac1N}\sum_{{\bs\alpha}\in{\rm roots}(\mathfrak{gl}_N)} c_{\bs\alpha} V_{\bs\alpha}(w)
+\ldots\,,
\ee
one finds, that $\tilde\Psi_\alpha(z)\Psi_\beta(w) =
\delta_{\alpha\beta}\mathbf{1}\cdot(z-w)^{\frac{1-N}{N}} + reg.$, i.e.
these fields look almost like fermions, except for the wrong power in the OPE. To fix this let us add an
extra scalar field $\phi(z)$, such that
\be
\phi(z)\phi(w)=-\frac1N\log(z-w)+\ldots
\ee
and define the new, the true fermionic, vertex operators
\be
\psi_{\alpha}(z)=e^{-i\phi(z)}\Psi_\alpha(z),\ \ \ \
\tilde\psi_{\alpha}(z)=e^{i\phi(z)}\tilde{\Psi}_\alpha(z),\ \ \ \ \alpha=1,\ldots,N
\ee
which have the canonical OPE (cf. with \rf{NOPE})
\be
\psi_\alpha(z)\tilde\psi_\beta(w)=\frac{\delta_{\alpha\beta}}{z-w}+reg.\\
\psi_\alpha(z)\psi_\beta(w)=reg.\ \ \ \
\tilde{\psi}_\alpha(z)\tilde\psi_\beta(w)=reg.
\ee
The rest is to understand, how to express the W-algebra generators in terms of these free fermions. One can easily write for the structure of the sum
\be
(z-w)^{-1/N}\Sum_{\alpha}\tilde\Psi_\alpha(z)\Psi_\alpha(w)=\frac{\mathbf{1}}{z-w}+\#\cdot(z-w)(\mathcal{L}_{-2} \mathbf{1})(w)+\\+
\#(z-w)^2\cdot(\mathcal{L}_{-1}\mathcal{L}_{-2}\mathbf{1})+\#(z-w)^2\cdot(\mathcal{W}_{-3}\mathbf{1})(w) +\ldots =\\=
\frac1{z-w}+\#\cdot(z-w)T(w)+\#\cdot(z-w)^2\pd T(w)+\#\cdot(z-w)^2 W(w)+...
\ee
with some coefficients (and where we have used obvious notations for the descendants). We do not need their exact numeric values at the moment, just the very
fact that only the unit operator $\mathbf{1}$ enters the r.h.s. of this OPE together with its descendants. Using additionally the OPE of the $U(1)$ factors
\be
(z-w)^{1/N}e^{i\phi(z)}e^{-i\phi(w)}=
\\
= :\exp\left(i(z-w)\pd\phi(w)+\frac12 i(z-w)^2\pd^2\phi(w)+\frac16(z-w)^3\pd^3\phi(w)\right):=
\\
=
1+(z-w)j(w)+\frac12(z-w)^2\pd j(w)+\frac16(z-w)^3\pd^2 j(w)+\\+\frac12(z-w)^2:j(w)^2:+\frac12(z-w)^3:j(w)\pd j(w):+ \frac16(z-w)^3 j(w)^3+\ldots
\ee
one can get
\be
\Sum_{\alpha}\tilde\psi_\alpha(z)\psi_\alpha(w)=\frac1{z-w}+j(w)+(z-w)\left(\#\cdot T(w)+\frac12 j(w)+\frac12 :j(w)^2:\right)+
\\+(z-w)^2\left(\#\cdot W(w)+\#\cdot j(w)T(w)+\frac16 \pd^2j(w)+\frac12 :j(w)\pd j(w):+\frac16 :j(w)^3:\right)+...\label{W0gen}
\ee
This formula states, how the standard W-generators can be expressed via the fermionic bilinears by some triangular transformation, and its symmetric form is equivalent to \rf{psipsi}, \rf{UjT}.

\setcounter{equation}0
\section{Vertex operators and Riemann-Hilbert problem}

\subsection{Vertex operators and monodromies}

Let us now turn to general construction of the monodromy vertex operator~\footnote{Notice, that
we have here only the conservation of the ``total charge'' $\sum_\alpha \sigma_\alpha +
\sum_\alpha \nu_\alpha = \sum_\alpha \theta_\alpha$, and apart of that their values are
arbitrary.}
\be
V_{\bs\nu}(t)\colon\mc{H}^{\bs\sigma}\to\mc{H}^{\bs\theta}
\ee
Actually one can define only the operator $V_{\bs\nu}(1)$ due to conformal Ward identity
\be
V_{\bs\nu}(t)=t^{-\Delta_{\bs\nu}}t^{L_0}V_{\bs\nu}(1)t^{-L_0}
\label{Vt}
\ee
and the operator $V_{\bs\nu}(1)$ is defined by the following three properties:
\begin{itemize}
\item $V_{\bs\nu}(1)$ is a (quasi)-group element, i.e. $$
    V_{\bs\nu}(1)\mc{H}^{\bs\sigma}\left(V_{\bs\nu}(1)\right)^{-1}\subseteq\mc{H}^{\bs\theta},\ \ \
 \left(V_{\bs\nu}(1)\right)^{-1}\mc{H}^{\bs\theta}V_{\bs\nu}(1)\subseteq\mc{H}^{\bs\sigma}$$
As we discussed already in sect.~\ref{ss:u1} this fact actually implies that all correlators of fermions in the presence of such an operator can be computed using the Wick theorem.
\item $\langle\bs\theta|V_{\bs\nu}(1)|\bs\sigma\rangle=1$, which is a kind of convenient normalization. Notice, however, that vertex operator is defined by the adjoint action only up to some diagonal factor
    $S=\exp(\bs\beta)$, $\bs\beta\in \mathfrak{h} \subset \mathfrak{gl}(N)$. In what follows we shall
    restore these diagonal factors when necessary.
\item  All two-fermionic correlators give the solution for the 3-point Riemann-Hilbert problem in the different regions
\be
\langle\bs\theta|V_{\bs\nu}(1){\tilde\psi}^{\bs\sigma}_\alpha(z)\psi^{\bs\sigma}_\beta(w)
|\bs\sigma\rangle=\mc K_{\alpha\beta}(z,w),\qquad |z|\leq 1$, $|w|\leq 1
\\
\langle\bs\theta|{\tilde\psi}^{\bs\theta}_{\dot\alpha}(z)\psi^{\bs\theta}_{\dot\beta}(w)V_{\bs\nu}(1)
|\bs\sigma\rangle=\mc K_{\dot\alpha\dot\beta}(z,w),\qquad |z|\geq 1$, $|w|\geq 1
\\
\langle\bs\theta|{\tilde\psi}^{\bs\theta}_{\dot\alpha}(z)V_{\bs\nu}(1)\psi^{\bs\sigma}_\beta(w)
|\bs\sigma\rangle=\mc K_{\dot\alpha\beta}(z,w),\qquad |z|\geq 1$, $|w|\leq 1
\\
-\langle\bs\theta|\psi^{\bs\theta}_{\dot\beta}(w)V_{\bs\nu}(1){\tilde\psi}^{\bs\sigma}_\alpha(z)
|\bs\sigma\rangle=\mc K_{\alpha\dot\beta}(z,w),\qquad |z|\leq 1$, $|w|\geq 1
\label{RH3K}
\ee
In terms of some matrix kernels $\mathcal{K}(z,w) = \mathcal{K}^{\bs\nu}(z,w)$,
where we have used $\{\alpha,\beta\}$ and $\{\dot\alpha,\dot\beta\}$ to denote matrix indices, corresponding to different bases, associated with the points $z=0$ and $z=\infty$ respectively.

\end{itemize}

By this moment the only claim is that this operator is uniquely defined by th properties listed above, and this
follows from the fact, that all matrix elements of the quasi-group $V_{\bs\nu}(1)$ element are given by certain determinants of the matrices with the entries, constructed from $\mathcal{K}(z,w)$. Existence of this operator is therefore obvious, since one can compute all its matrix elements using the Wick theorem.

Now, we would like to specify the kernels $\mathcal{K}(z,w)$ first by their monodromy properties. We associate
the basis at $z=0$ with the eigenvectors of $M_0\sim e^{2\pi i\bs\sigma}$, while the basis at $z=\infty$ with the eigenvectors of $M_\infty\sim e^{2\pi i\bs\theta}$ (only the conjugacy classes of these two matrices are fixed, and
certainly in general $[M_0,M_\infty]\neq 0$). We propose an explicit form of the kernel
\be
\mc K_{\alpha\beta}(z,w)=\frac{[\phi(z)\phi(w)^{-1}]_{\alpha\beta}}{z-w}
\label{Kphi}
\ee
given in terms of the solution to the linear system
\be
\frac{d}{dz}\phi(z)=\phi(z)\left(\frac{A_0}z+\frac{A_1}{z-1}\right)=\phi(z)A(z)
\label{lin_system}
\ee
with $A_0\sim\bs\sigma$, $A_1\sim\bs\nu$, $A_\infty\sim\bs\theta$ and
prescribed monodromies
\eqs{
\gamma_0\colon& \phi_{\alpha i}(z)\mapsto \sum_{\beta} (M_0)_{\alpha\beta}\phi(z)_{\beta i}\\
\gamma_\infty\colon& \phi_{\alpha i}(z)\mapsto \sum_{\beta} (M_\infty)_{\alpha\beta}\phi(z)_{\beta i}
}
also implying monodromy around $z=1$, i.e. $\gamma_1: \phi_{\alpha i}(z)\mapsto \sum_{\beta} (M_1)_{\alpha\beta}\phi(z)_{\beta i}$, with $M_1\sim e^{2\pi i\bs\nu}$ and $M_0M_1M_\infty=1$.
Solutions for a linear system \rf{lin_system} can be expressed themselves in terms of a
fermionic correlators, namely
\eqs{\phi_{\alpha\gamma}(z)=&z\cdot\langle\bs\theta|V_{\bs\nu}(1)\tilde\psi_\alpha(z)|-\bs1_\gamma,\bs\sigma\rangle\\
\phi^{-1}_{\gamma\beta}(z)=&z\cdot\langle\bs\theta|V_{\bs\nu}(1)\psi_\beta(z)|\bs1_\gamma,\bs\sigma\rangle
\label{phi1ferm}}
for some fixed normalization at $z\to 0$. We are going to prove in next section, that definitions \rf{Kphi} are indeed self-consistent and also consistent with \rf{phi1ferm}, which follows from the generalized Hirota bilinear relations, satisfied by the monodromy vertex operators.

Actually, we have four different matrix kernels \rf{Kphi} with the indices $\alpha\beta$, $\alpha\dot\beta$, $\dot\alpha\beta$, $\dot\alpha\dot\beta$, corresponding to all possible combinations of different regions.
When we change from one region to another one, then we have to change the
basis of solutions, and this transition can be given by some matrix $C_{\dot\alpha}^\alpha$.

The expansion of these kernels, e.g. for $0<z,w<1$
\be
\mathcal{K}_{\alpha\beta}(z,w)={\delta_{\alpha\beta}\over z-w} +
\Sum_{p,q>0}\langle{\bs\theta}|V_{\bs\nu}(1)\tilde{\psi}{^\alpha}_{-p}\psi^\beta_{-q}|\bs\sigma\rangle z^{p-\frac12+\sigma_\alpha}w^{q-\frac12-\sigma_\beta} =
\\
={\delta_{\alpha\beta}\over z-w} +
\Sum_{p,q>0}K_{pq}^{\alpha\beta}z^{p-\frac12+\sigma_\alpha}w^{q-\frac12-\sigma_\beta}
\label{maelK1}
\ee
or at $z,w>1$
\be
\mathcal{K}_{\dot{\alpha}\dot{\beta}}(z,w)={\delta_{\dot{\alpha}\dot{\beta}}\over z-w} +
\Sum_{p,q>0}\langle\bs\theta|\tilde{\psi}^{\dot\alpha}_{p}\psi^{\dot\beta}_{q}V_{\bs\nu}(1)|{\bs\sigma}\rangle
z^{-p-\frac12+\theta_{\dot\alpha}}w^{-q-\frac12-\theta_{\dot\beta}}=
\\
={\delta_{\dot{\alpha}\dot{\beta}}\over z-w} +
\Sum_{p,q>0}\tilde{K}_{pq}^{\dot\alpha\dot\beta}z^{-p-\frac12+\theta_{\dot\alpha}}w^{-q-\frac12-\theta_{\dot\beta}}
\label{maelK2}
\ee
give the corresponding matrix elements for the fermionic modes. The corresponding matrix elements ($K_{p_\alpha,q_\beta}^{\alpha\beta}$ and $\tilde{K}_{p_\alpha,q_\beta}^{\alpha\beta}$) are in fact defined
up to the factors $s_\alpha s_\beta^{-1}$ which comes from the ambiguity in normalization of the vertex operator. For any three monodromy matrices $M_0 M_1 M_\infty=1$ one can fix all their
invariant functions (e.g. traces) and diagonalize $M_\infty$, but then one possible transformation survives: a simultaneous conjugation
\be
M_i\mapsto S^{-1}M_i S
\ee
by diagonal $S=\mathrm{diag}(s_1,\ldots,s_N)$. This gives vertex operators, actually different by $\bs s^{\bs J_0}$ factor with corresponding multiplicative renormalization of their matrix elements.

For special vertex operators with
$\bs\nu = \nu N \mathbf{e}_j$ these matrix elements can be expressed in terms of the products \rf{fprod}, for example
\be
K_{p_\alpha,q_\beta}^{\alpha\beta} =
\langle\bs\theta|V_\nu(1)|p_\alpha,q_\beta;\bs\sigma\rangle =
\langle{\bs\theta}|V_\nu(1)\tilde{\psi}{^\alpha}_{-p_\alpha}\psi^\beta_{-q_\beta}|\bs\sigma\rangle =
\\
=\frac{1}{p_\alpha+q_\beta-\sigma_\alpha+\sigma_\beta}
f_{1,\alpha}(\bs\sigma,\bs\theta+\nu,p_\alpha)f_{2,\beta}(\bs\sigma,\bs\theta+\nu,q_\beta)
\\
\tilde{K}_{p_\alpha,q_\beta}^{\alpha\beta} = \langle p_\alpha,q_\beta;\bs\theta|V_\nu(1)|\bs\sigma\rangle =
\langle\bs\theta|\tilde{\psi}^{\alpha}_{p_\alpha}\psi^{\beta}_{q_\beta}V_\nu(1)|{\bs\sigma}\rangle =
\\
= -\frac{1}{p_\alpha+q_\beta-\theta_\alpha+\theta_\beta}
f_{1,\alpha}(\bs\theta+\nu,\bs\sigma,p_\alpha)f_{2,\beta}(\bs\theta+\nu,\bs\sigma,q_\beta)
\ee
We shall return to discussion of special case below in sect.~\ref{ss:hyper}.

The general formula for $2n$-point fermionic correlator is given by the Wick formula
\eqs{
\langle\bs\theta|\prod_{\dot\alpha=1}^N\prod_{i=1}^{d'_{\dot\alpha}}
{\tilde\psi}^{\bs\theta}_{\dot\alpha}(z_{\dot\alpha,i})
\psi^{\bs\theta}_{\dot\alpha}(w_{\dot\alpha,i}) V_{\bs\nu}(1)\prod_{\alpha=1}^N\prod_{i=1}^{d_\alpha}{\tilde\psi}^{\bs\sigma}_{\alpha}(z_{\alpha,i})
\psi^{\bs\sigma}_{\alpha}(w_{\alpha,i})|\bs\sigma\rangle=\\
=\langle\bs\theta|V_{\bs\nu}(1)|\bs\sigma\rangle\cdot\det\bpm\mc K_{\alpha\beta}
(z_{\alpha,i},w_{\beta,j})&
\mc K_{\alpha\dot\beta}(z_{\alpha,i},w_{\dot\beta,j})\\\mc K_{\dot\alpha\beta}(z_{\dot\alpha,i},w_{\beta,j})&
\mc K_{\dot\alpha\dot\beta}(z_{\dot\alpha,i},w_{\dot\beta,j})\epm\label{ferm_corr}
}
On the punctured unit circle $|z|=|w|=1$, $z\neq 1$, $w\neq 1$
one has
\eqs{
\mc K_{\dot\alpha\beta}(z,w)=&\Sum_\alpha C^\alpha_{\dot\alpha}\mc K_{\alpha\beta}(z,w)\\
\mc K_{\alpha\dot\beta}(z,w)=&\Sum_\beta\mc K_{\alpha\beta}(z,w)\left(C^{-1}\right)^\beta_{\dot\beta}
}
It follows then from \rf{ferm_corr}, that there are two operator identities
\eqs{
{\tilde\psi}^{\bs\theta}_{\dot\alpha}(z)V_{\bs\nu}(1)=&V_{\bs\nu}(1)\Sum_\alpha C^\alpha_{\dot\alpha}{\tilde\psi}^{\bs\sigma}_\alpha(z)\\
\psi^{\bs\theta}_{\dot\alpha}(z)V_{\bs\nu}(1)=&V_{\bs\nu}(1)\Sum_\alpha \psi^{\bs\sigma}_\alpha(z)(C^{-1})^\alpha_{\dot\alpha}
\label{jump}
}
Actually, these identities are enough to define the operator $V_{\bs\nu}(1)$. The simplest quantity to compute is
\eq{V_{\bs\nu}(1)\psi^{\bs\sigma}_{\alpha,r}V_{\bs\nu}(1)^{-1}=\Oint_{|z|=1}\frac{dz}{2\pi i}z^{r-\frac12+\sigma_\alpha}
V_{\bs\nu}(1)\psi^{\bs\sigma}_{\alpha}(z)V_{\bs\nu}(1)^{-1}
}
Using \rf{jump} one can rewrite this equivalently as
\eq{V_{\bs\nu}(1)\psi^{\bs\sigma}_{\alpha,r}V_{\bs\nu}(1)^{-1}=\Sum_{\beta}C^\beta_\alpha
\Oint_{|z|=1}\frac{dz}{2\pi i}z^{r-\frac12+\sigma_\alpha}
\psi_\beta^{\bs\theta}(z)=\\=\Sum_{\beta,s}C^\beta_\alpha\Oint_{|z|=1}\frac{dz}{2\pi i}z^{r-s-1+\sigma_\alpha-\theta_\beta}
\psi_{\beta,s}^{\bs\theta}=\Sum_{\beta,s}C^\beta_\alpha\Int_0^{2\pi}\frac{d\phi}{2\pi}e^{2\pi i(r-s+\sigma_\alpha-\theta_\beta)\phi}
\psi^{\bs\theta}_{\beta,s}=\\=
\Sum_{\beta,s}C_\alpha^\beta \frac{-i(e^{2\pi i(\sigma_\alpha-\theta_\beta)}-1)}{r-s+\sigma_\alpha-\theta_\beta}\psi^{\bs\theta}_{\beta,s}
}
In principle, this formula includes all possible information about $V_{\bs\nu}(t)$.
Now it is easy to prove
\begin{theorem}
$V_{\bs\nu}(t)$ is a primary field of the conformal $W_N\oplus H$ algebra with the highest weights $u_k(\bs\nu)$.
\end{theorem}
{\bf Proof:} First we notice that due to \rf{jump} and to the definitions \rf{Vt}, \rf{Wdef} one has
\be
U_k^{\bs\theta}(z)V_{\bs\nu}(t)=V_{\bs\nu}(t)U_k^{\bs\sigma}(z)
\ee
in the region $|z|=t$, $z\neq t$. This means that $U_k(z)$ are actually single-valued operators (with trivial monodromies).
Actually, we have already proved in Theorem \ref{thm:highest_weight} that states $|\bs\sigma\rangle$ are highest weight vectors, so
\be
\langle\bs\theta|\ldots U_k(z)|\bs\sigma\rangle=\left(\frac{u_k(\bs\sigma)}{z^k}+\text{less singular}\right)\langle\bs\theta|\ldots|\bs\sigma\rangle
\ee
and, since \rf{lin_system} is symmetric under the permutation of the singular points, one can also conclude, that for a different point
\be
\langle\bs\theta|\ldots U_k(z)V_{\bs\nu}(t)\ldots|\bs\sigma\rangle=\left(\frac{u_k(\bs\nu)}{(z-t)^k}+\text{less singular}\right)
\langle\bs\theta|\ldots V_{\bs\nu}(t)\ldots|\bs\sigma\rangle
\ee
so $(\mc U_{k,n>0}V_{\bs\nu})(t)=0$, and it means, that $V_{\bs\nu}(t)$ is just a primary field. \hfill $\square$

\subsection{Generalized Hirota relations}

Now consider \emph{any} operator $\mc{O}$  with linear adjoint action on fermions
\be
\label{adjgl}
\mc{O}^{-1}\psi_{\alpha,r}\mc{O}=\Sum_{s,\beta}R^{\mc{O}}_{r\alpha,s\beta}\psi_{\beta,s}\,,\ \ \
\mc{O}^{-1}\tilde\psi_{\alpha,-r}\mc{O}=\Sum_{s,\beta}\tilde\psi_{\beta,-s}(R^{\mc{O}})^{-1}_{s\beta,r\alpha}
\ee
which is generally a relabeling of a $GL(\infty)$ transformation for a single fermion.
It leads to a standard statement of commutativity of two operators in $\mathcal{H}\otimes \mathcal{H}$
\be
\label{Hirop}
\mc{O}\otimes\mc{O}\Sum_{r,\alpha}\psi_{\alpha,-r}\otimes\tilde\psi_{\alpha,r}=
\Sum_{r,\alpha}\psi_{\alpha,-r}\otimes\tilde\psi_{\alpha,r}\mc{O}\otimes\mc{O}
\ee
which is an operator form of the bilinear Hirota relation \cite{Hirota,AZ}.

Let us now point out, that we have already introduced by \rf{jump}
a particular subclass of general transformations \rf{adjgl}
\be
\label{mongl}
V^{-1}\psi_{\dot\alpha}(z)V=\Sum_{\alpha}(C^{-1})^\alpha_{\dot\alpha}\psi_\alpha(z)\,,\ \ \
V^{-1}\tilde\psi_{\dot\alpha}(z)V=\Sum_{\alpha}C^\alpha_{\dot\alpha}\tilde\psi_\alpha(z)
\ee
where $C$ and $C^{-1}$ can be now interpreted as monodromy matrices: one can consider \rf{mongl} as a linear relation between two analytic continuations
of the fermionic fields at $|z|=1$ towards $z\to\infty$ and $z\to 0$, preserving the OPE $\tilde\psi(z)_{\alpha}(z)\psi_\beta(z') = {\delta_{\alpha\beta}\over z-z'} + \ldots$.
An immediate consequence of \rf{mongl} is
\begin{theorem}\label{thm:bilinear_operator}
The Fourier modes of the bilinear operators
\be
{\mc I}(z)=\Sum_{\alpha}\psi_\alpha(z)\otimes\tilde{\psi}_\alpha(z)=\Sum_{k\in\mathbb Z}\frac{{\mc I}_k}{z^{k+1}}\\
{\mc I}^\dagger(z)=\Sum_\alpha\tilde{\psi}_\alpha(z)\otimes\psi_\alpha(z)=\Sum_{k\in\mathbb Z}\frac{{\mc I}^\dagger_k}{z^{k+1}}\label{Xmodes}
\ee
commute with $V_{\bs\nu}(t)\otimes V_{\bs\nu}(t)$ in the sense
\be
{\mc I}_k^{\bs\theta}\cdot V_{\bs\nu}(t)\otimes V_{\bs\nu}(t) =V_{\bs\nu}(t)\otimes V_{\bs\nu}(t)\cdot {\mc I}_k^{\bs\sigma}\\
{{\mc I}^\dagger}_k^{\bs\theta}\cdot V_{\bs\nu}(t)\otimes V_{\bs\nu}(t) =V_{\bs\nu}(t)\otimes V_{\bs\nu}(t)\cdot {\mc I}_k^{\bs\sigma}
\ee
\end{theorem}
{\bf Proof:}
First we notice that
\eq{{\mc I}^{\bs\theta}(z)\cdot V_{\bs\nu}(t)\otimes V_{\bs\nu}(t)=V_{\bs\nu}(t)\otimes V_{\bs\nu}(t)\cdot {\mc I}^{\bs\sigma}(z)
}
holds at $|z|=t$, $z\neq t$, due to \rf{jump}
\eq{
\Sum_\alpha\psi^{\bs\theta}_\alpha(z)\otimes{\tilde\psi}^{\bs\theta}_\alpha(z)\cdot V_{\bs\nu}(t)\otimes V_{\bs\nu}(t)=
V_{\bs\nu}(t)\otimes V_{\bs\nu}(t)\Sum_{\alpha,\dot\beta,\dot\gamma}
(C^{-1})^{\dot\beta}_\alpha C^{\dot\gamma}_\alpha\psi^{\bs\sigma}_{\dot\beta}(z)\otimes
{\tilde\psi}^{\bs\sigma}_{\dot\gamma}(z)=\\=
V_{\bs\nu}(t)\otimes V_{\bs\nu}(t)\Sum_{\dot\beta}\psi^{\bs\sigma}_{\dot\beta}(z)\otimes{\tilde\psi}^{\bs\sigma}_{\dot\beta}(z)
}
To continue this equality to $z=t$ one has just to check that
${\mc I}^{\bs\theta}(z)\cdot V_{\bs\nu}(t)\otimes V_{\bs\nu}(t)$ is regular. Due to the symmetry of \rf{lin_system} this is the same as to check that ${\mc I}^{\bs\sigma}(z)\cdot|\bs\sigma\rangle\otimes|\bs\sigma\rangle$ is regular. Since,
\eq{{\mc I}^{\bs\sigma}(z)\cdot|\bs\sigma\rangle\otimes|\bs\sigma\rangle=\Sum_\alpha\Sum_{n<0}\frac{\psi^{\bs\sigma}_{\alpha,n}|\bs\sigma\rangle}{z^{n+\frac12
+\sigma_\alpha}}\otimes\Sum_{m<0}\frac{{\tilde\psi}^{\bs\sigma}_{\alpha,m}|\bs\sigma\rangle}{z^{m+\frac12-\sigma_\alpha}}=\\=\Sum_\alpha
\psi_{\alpha,-\frac12}^{\bs\sigma}|\bs\sigma\rangle\otimes{\tilde\psi}^{\bs\sigma}_{\alpha,-\frac12}|\bs\sigma\rangle+O(z)
}
this expression is regular, this completes the proof.
\hfill $\square$

Let us notice that we have also got the equalities
\eq{{\mc I}^{\bs\sigma}_{k\geq0}\cdot|\bs\sigma\rangle\otimes|\bs\sigma\rangle=0,\qquad
{{\mc I}^\dagger}^{\bs\sigma}_{k\geq0}\cdot|\bs\sigma\rangle\otimes|\bs\sigma\rangle=0\label{Xhighest}
}
while, for example
\be
\mathcal{I}^\dagger_{-1}|{\bs\theta}\rangle\otimes|{\bs\theta}\rangle = \Sum_{\alpha}\tilde\psi_{\alpha,-1/2}\otimes\psi_{\alpha,-1/2}|{\bs\theta}\rangle\otimes|{\bs\theta}\rangle = \Sum_{\alpha}|\mathbf{1}_\alpha,{\bs\theta}\rangle\otimes|-\mathbf{1}_\alpha,{\bs\theta}\rangle
\\
\mathcal{I}_{-1}|{\bs\theta}\rangle\otimes|{\bs\theta}\rangle = \Sum_{\alpha}\psi_{\alpha,-1/2}\otimes\tilde\psi_{\alpha,-1/2}|{\bs\theta}\rangle\otimes|{\bs\theta}\rangle = \Sum_{\alpha}|-\mathbf{1}_\alpha,{\bs\theta}\rangle\otimes|\mathbf{1}_\alpha,{\bs\theta}\rangle
\label{Im1}
\ee
but
\be
\langle\bs\theta|\otimes\langle\bs\theta|\cdot {\mc I}^\dagger_{-1}=
\langle\bs\theta|\otimes\langle\bs\theta|\cdot {\mc I}_{-1}=0
\label{Im1bra}
\ee
We shall see below, that existence of extra bilinear operator relations lead actually to the infinite number of Hirota-like equations for the $\tau$-function.

Let us also notice that operator $t^{L_0}$ belongs to the quasigroup, but it does not commute with $\mc I_k$:
\be
t^{L_0}\mc I(z)t^{-L_0}=t\mc I(tz)
\ee
which means that $t^{L_0}\mc I_kt^{-L_0}=t^{-k}$. So, in principle, vertex operator can contain some factors $t_i^{L_0}$, but in such a combination with $\prod t_i=1$.

Now we are ready to prove, that the correlation functions \rf{RH3K} (and in fact any correlation function
$\langle\bs\theta_{\infty}|{\mc O}\tilde\psi_\alpha(z)\psi_\beta(w)|\bs\theta_0\rangle=
\langle\bs\theta_{\infty}|V_{\bs\theta_{n-2}}(t_{n-2})\ldots V_{\bs\theta_1}(t_1)\tilde\psi_\alpha(z)\psi_\beta(w)|\bs\theta_0\rangle$
with two fermions)
can be decomposed into two correlation functions with a single fermion insertion.
In addition to \rf{Im1}, \rf{Im1bra} one has to compute
commutator of this operator with $\tilde{\psi}\otimes\psi$ using the contour integral representation
\eq{
\left[{\mc I}_{-1},\tilde\psi_\alpha(z)\otimes\psi_\beta(w)\right]=\left(\Oint_z+\Oint_w\right)\frac{dx}{x}
\Sum_{\gamma}\psi_\gamma(x)\otimes
\tilde\psi_\gamma(x)\cdot\tilde\psi_\alpha(z)\otimes\psi_\beta(w)=\\=
\Oint_z\frac{dx}{x}\Sum_\gamma\frac{\delta_{\gamma\alpha}}{x-z}\otimes\tilde\psi_\gamma(x)\psi_\beta(w)+
\Oint_w\frac{dx}{x}\Sum_\gamma\psi_\gamma(x)\tilde\psi_\alpha(z)\otimes\frac{\delta_{\gamma\beta}}{x-w}=
\\
= \frac1z\cdot 1\otimes\tilde\psi_\alpha(z)\psi_\beta(w) +
\frac1w\cdot \psi_\beta(w)\tilde\psi_\alpha(z)\otimes 1
}
Inserting this operator identity inside the correlation functions, and using \rf{Im1}, \rf{Im1bra} we get
\eq{0=\langle\bs\theta_\infty|\otimes\langle\bs\theta_\infty|\cdot {\mc I}_{-1}\cdot\mc O\otimes\mc O\cdot\tilde\psi_\alpha(z)\otimes\psi_\beta(w)
\cdot|\bs\theta_0\rangle\otimes|\bs\theta_0\rangle=\\=
\langle\bs\theta_\infty|\otimes\langle\bs\theta_\infty|\cdot\mc O\otimes\mc O
\cdot\tilde\psi_\alpha(z)\otimes\psi_\beta(w)\Sum_\gamma|-\bs 1_\gamma,\bs\theta_0\rangle\otimes|\bs 1_\gamma,\bs\theta_0\rangle +
\\
+\left(\frac1z-\frac1w\right)
\langle\bs\theta_\infty|{\mc O}|\bs\theta_0\rangle \cdot \langle\bs\theta_\infty|{\mc O}\tilde\psi_\alpha(z)\psi_\beta(w)|\bs\theta_0\rangle
\label{equality1}
}
The first term in the r.h.s. is equal to the bilinear combination of the correlation functions
with a single fermion insertion, so one gets finally
\eq{
\langle\bs\theta_\infty|\mc O\tilde\psi_\alpha(z)\psi_\beta(w)|\bs\theta_0\rangle\langle\bs\theta_\infty|\mc O|\bs\theta_0\rangle=
\\=
\frac{zw}{z-w}\Sum_{\gamma}\langle\bs\theta_\infty|\mc O \tilde\psi_\alpha(z)|-\bs 1_\gamma,\bs\theta_0\rangle\langle\bs\theta_\infty|\mc O
\psi_\beta(w)|\bs 1_\gamma,\bs\theta_0\rangle\label{factorization}
}
which for $\mc O = V_{\bs\nu}(1)$ gives the relation between \rf{phi1ferm} and \rf{Kphi}. Substituting
here the OPE $\tilde\psi_\alpha(z)\psi_\beta(w)=\frac{\delta_{\alpha\beta}}{z-w}+\text{reg.}$ and taking
residue at $z\to w$ one also proves that matrices in \rf{phi1ferm} are indeed inverse to each other.

\subsection{Riemann-Hilbert problem: hypergeometric example
\label{ss:hyper}}

A hypergeometric solution to the Riemann-Hilbert problem with three singular points at $z=0,1,\infty$ can be given by the following formulas
\begin{align*}
\phi(z)=&
 \begin{pmatrix}
   z^\beta \mathcal F(\alpha,\beta,\nu|z) &  -z^{1+\beta}C(\alpha,\beta,\nu)\mathcal F(\alpha,1+\beta,\nu|z)
\\
  -z^{1-\beta}C(\alpha,-\beta,\nu)\mathcal F(\alpha,1-\beta,\nu|z) &  z^{-\beta}\mathcal F(\alpha,-\beta,\nu|z)
 \end{pmatrix},
\\[1mm]
 \phi^{-1}(z)=&
 \begin{pmatrix}
  z^{-\beta} \mathcal F(-\alpha,-\beta,-\nu|z) & z^{1+\beta}C(\alpha,\beta,\nu)\mathcal F(\alpha,1+\beta,-\nu|z)
\\
  z^{1-\beta}C(\alpha,-\beta,\nu)\mathcal F(\alpha,1-\beta,-\nu|z) &  z^\beta\mathcal F(\alpha,\beta,-\nu|z)
 \end{pmatrix}
\end{align*}
where we have introduced
$ \mathcal F(\alpha,\beta,\nu|z)={}_2F_1\bigl[\substack{-\alpha+\beta-\nu,\,\alpha+\beta-\nu \\ 2\beta}\big|z\bigr]$
for a standard hypergeometric function and the constant $C(\alpha,\beta,\nu)=\frac{(-\alpha-\beta+\nu)(\alpha -\beta+\nu)}{2\beta(2\beta+1)}$.

These formulas give solution to the linear system \rf{lin_system}
with the residues in the following conjugacy classes:
\eq{
A_0\sim{\bs\theta}_0={\bs\sigma}=\mathrm{diag}(\beta,-\beta),\qquad A_\infty\sim{\bs\theta}_\infty={\bs\theta}=\mathrm{diag}(\alpha,-\alpha)
\\
 A_1\sim{\bs\theta}_1={\bs\nu}=\mathrm{diag}(2\nu,0)
}
According to \rf{RH3K}, \rf{Kphi}
\eq{
\langle{\bs\theta}| V_{\bs\nu}\tilde\psi_\alpha(z)\psi_\beta(w)|{\bs\sigma}\rangle
=\frac{[\phi(z)\phi(w)^{-1}]_{\alpha\beta}}{z-w}
}
It means, for example, that in order to study the matrix elements with $\psi_1$, $\tilde \psi_1$ one needs to consider the function
\eq{
\widehat{\mathcal{K}}_{11}(z,w)=z^{-\beta}w^{\beta}[\phi(z)\phi(w)^{-1}]_{11}=\mathcal F(\alpha,-\beta,-\nu|z)\mathcal F(\alpha,\beta,\nu|w)-\\-
\frac{\prod\limits_{\epsilon,\epsilon'=\pm1}(\epsilon\alpha+\epsilon'\beta+\nu)}{4\beta^2(4\beta^2-1)}zw
\mathcal F(\alpha,1-\beta,-\nu|z)\mathcal F(\alpha,1+\beta,\nu|w)
}
Already the simplest fact, that $\widehat{\mathcal{K}}_{11}(z,z)=1$ becomes a non-trivial bilinear relation for the hypergeometric function. However, our claim is much stronger: this function
is almost as nice as \rf{identity1_} since its expansion \rf{maelK1} is given by
\eq{
\mathcal{K}(z,w)_{11} = \frac{\widehat{\mathcal{K}}_{11}(z,w)}{z-w}=\frac1{z-w}-
\\
-\Sum_{a,b=0}^\infty\frac{2\beta(\alpha-\beta+\nu)_{a+1}(-\alpha+\beta+\nu)_{a+1}(-\alpha+\beta-\nu)_{b+1}
(\alpha+\beta-\nu)_{b+1}}{(a+b+1)(-\alpha+\beta-\nu)(\alpha+\beta-\nu)a!b!(2\beta)_{b+1}(-2\beta)_{a+1}}z^bw^a
}
and it is indeed a generation function of the matrix elements we are interested in.
One can substitute here $a=q-\frac12$, $b=p-\frac12$
\eq{\label{element1}
\langle(\alpha,-\alpha)|V_{(2\nu,0)}(1)\psi_{1,-p}\tilde\psi_{1,-q}|(\beta,-\beta\rangle=\\=
\frac{2\beta(\alpha+\beta-\nu)_{q+\frac12}(-\alpha+\beta-\nu)_{q+\frac12}(\alpha-\beta+\nu)_{p+\frac12}
(-\alpha-\beta+\nu)_{p+\frac12}}{(p+q)(-\alpha+\beta-\nu)(\alpha+\beta-\nu)(p-\frac12)!(q-\frac12)!(2\beta)_{q+\frac12}(-2\beta)_{p+\frac12}}
}
and compare this formula with \rf{UNdet}
\eq{\frac{f_{1,1}(\bs\theta,\bs\theta',p)f_{2,1}(\bs\theta,\bs\theta',q)}{p+q}=
\frac1{(p-\frac12)!}\prod_\beta\frac{(\theta_\beta'-\theta_1)_{p+\frac12}
}{\sqrt{\theta_\beta'-\theta_1}}\prod_{\beta\neq1}\frac{\sqrt{\theta_\beta-\theta_1}}{
(\theta_\beta-\theta_1)_{p+\frac12}}\times\\\times
\frac1{(q-\frac12)!}\prod_\beta\frac{(\theta_1-\theta_\beta')_{q+\frac12}
}{\sqrt{\theta_1-\theta_\beta'}}\prod_{\beta\neq1}\frac{\sqrt{\theta_1-\theta_\beta}}{
(\theta_1-\theta_\beta)_{q+\frac12}}\times\frac1{p+q}=\\=
\frac{(\theta_1-\theta_2)(-\theta_1+\theta_1')_{p+\frac12}(-\theta_1+\theta_2')_{p+\frac12}(\theta_1-\theta_1')_{q+\frac12}
(\theta_1-\theta_2')_{q+\frac12}}{(p+q)(-\theta_1+\theta_1')(\theta_1-\theta_2')(p-\frac12)!(q-\frac12)!
(\theta_2-\theta_1)_{p+\frac12}(\theta_2-\theta_1)_{q+\frac12}}\label{element2}
}
It is easy to see, that after the appropriate identification
\eq{
\theta_1=\beta,\,\theta_2=-\beta,\,\theta_1'=\alpha+\nu,\,\theta_2'=-\alpha+\nu
}
the r.h.s.'s in two last formulas coincide exactly.

In addition to the hypergeometric case another explicit example can be provided by the exact conformal blocks, considered in \cite{GMtw}. We are planning to consider it in detail elsewhere.

\section{Isomonodromic tau-functions and Fredholm determinants
\label{ss:tauiso}}

\subsection{Isomonodromic tau-function}

First we need to prove the simple
\begin{lemma} Monodromies of $\psi_\beta(w)$ and $\tilde{\psi}_\alpha(z)$ in the matrix elements
\eq{\langle\bs Y',\bs n',\bs\theta|V_{\bs\nu}(1){\tilde\psi}^{\bs\sigma}_\alpha(z)\psi^{\bs\sigma}_\beta(w)|\bs Y,\bs n,\bs\sigma\rangle
}
do not depend on $\bs n, \bs Y, \bs n', \bs Y'$.\label{thm:monodromy}
\end{lemma}
{\bf Proof:} All these matrix elements can be obtained from \rf{ferm_corr} by certain contour integration, producing fermionic modes from the fermionic fields. However, in
\rf{ferm_corr} due to the Wick theorem factorization, all contributions have the factorized form $\mc K_{\alpha\gamma}(z,\bullet)\times\ldots$,
where all other factors do not depend at all on $z$, so that all monodromies comes from a single kernel $\mc K$. \hfill $\square$

Now it is easy to prove
\begin{theorem}
Solution of the linear problem with $n$ marked points is given by\par\noindent $(z-w)\mathfrak{K}_{\alpha\beta}(z,w)$ with
\eq{\mathfrak{K}_{\alpha\beta}(z,w)=\frac{\langle\bs\theta_{\infty}|V_{\bs\theta_{n-2}}(t_{n-2}) \ldots V_{\bs\theta_1}(t_1){\tilde\psi}^{\bs\theta_0}_
\alpha(z)\psi^{\bs\theta_0}_\beta(w)|\bs\theta_0\rangle}{\langle\bs\theta_{\infty}| V_{\bs\theta_{n-2}}(t_{n-2}) \ldots
V_{\bs\theta_1}(t_1)|\bs\theta_0\rangle
\label{solution}}
}
whereas its isomonodromic tau-function is defined by
\eq{\tau(t_1,\ldots,t_{n-2})=\langle\bs\theta_{\infty}| V_{\bs\theta_{n-2}}(t_{n-2}) \ldots
V_{\bs\theta_1}(t_1)|\bs\theta_0\rangle\label{tau_corr}
}
\end{theorem}
{\bf Proof:}
First, insert resolutions of unity between each
two (radially-ordered) vertex operators, e.g.
\eq{\tau\cdot \mathfrak{K}_{\alpha\beta}(z,w)=\Sum_{\{\bs  Y_1,\bs m_i\}}\langle\bs\theta_{\infty}|V_{\bs\theta_{n-2}}(t_{n-2})
|\bs Y_{n-3},\bs m_{n-3},\bs\sigma_{n-3}\rangle\langle\bs Y_{n-3},\bs m_{n-3},\bs\sigma_{n-3}|\times\\\times
\ldots\times \langle\bs Y_2,\bs m_2,\bs\sigma_2|V_{\bs\theta_2}(t_2)|\bs Y_1,\bs m_1,\bs\sigma_1\rangle
\langle\bs Y_1,\bs m_1,\bs\sigma_1|V_{\bs\theta_1}(t_1){\tilde\psi}^{\bs\theta_0}_
\alpha(z)\psi^{\bs\theta_0}_\beta(w)|\bs\theta_0\rangle
\label{expK}
}
for $0<|z|,|w|<|t_1|$ and similarly in the other regions.
Due to Lemma~\ref{thm:monodromy} the monodromies of
the fermionic fields do not depend on the intermediate states, but only on the vertex operators and the set of charges $\bs\sigma$'s~\footnote{In addition to $(n-3)$ time parameters ($\{t_1,\ldots,t_n\}$ modulo M\"obius transformation, which always allow to fix three of them to $0,1,\infty$) and $n$ sets of W-charges $\{{\bs\theta}_j\}$ the isomonodromic tau-function depends upon the charges $\{{\bs\sigma}_k\}\in (\mathbb{R}/\mathbb{Z})^{N-1}$, $k=1,\ldots,n-3$ in the intermediate channels and their duals $\{{\bs\beta}_k\}$, which we had already discussed in the context of ambiguity in normalization of the vertex operators and their matrix elements.}, therefore it is enough
to reduce the problem of computation of all monodromies to the collection of corresponding three-point problems
with different vertex operators $V_{\bs\theta_j}(t_j)$ inserted. So we have proven that $(z-w)\mathfrak{K}_{\alpha\beta}(z,w) = \left[\Phi(z)\Phi^{-1}(w)\right]_{\alpha\beta}$ (to cancel extra singularity in \rf{solution}), actually gives a solution to the multi-point Riemann-Hilbert problem.

In order to prove \rf{tau_corr} consider
\eq{
\Sum_{\alpha}\tilde{\psi}_\alpha(z+\frac t2)\psi_\alpha(z-\frac t2)=\frac{N}{t}+J(z)+t U_2(z)+\ldots
}
so that
\eq{t\ \Tr \mathfrak{K}(z+\frac t2,z-\frac t2)=\Tr\ \Phi(z+\frac t2)\Phi(z-\frac t2)^{-1}=\\=
N+t\frac{\langle\bs\theta_{\infty}|V_{\bs\theta_{n-2}}(t_{n-2}) \ldots V_{\bs\theta_1}(t_1)
J(z)|\bs\theta_0\rangle}{\langle\bs\theta_{\infty}| V_{\bs\theta_{n-2}}(t_{n-2}) \ldots
V_{\bs\theta_1}(t_1)|\bs\theta_0\rangle}+t^2\frac{\langle\bs\theta_{\infty}|V_{\bs\theta_{n-2}}(t_{n-2}) \ldots V_{\bs\theta_1}(t_1)
U_2(z)|\bs\theta_0\rangle}{\langle\bs\theta_{\infty}| V_{\bs\theta_{n-2}}(t_{n-2}) \ldots
V_{\bs\theta_1}(t_1)|\bs\theta_0\rangle}+\ldots
}
where from \rf{UjT} and the conformal Ward identities
\eq{
\frac{\langle\bs\theta_{\infty}|V_{\bs\theta_{n-2}}(t_{n-2}) \ldots V_{\bs\theta_1}(t_1)
U_2(z)|\bs\theta_0\rangle}{\langle\bs\theta_{\infty}| V_{\bs\theta_{n-2}}(t_{n-2}) \ldots
V_{\bs\theta_1}(t_1)|\bs\theta_0\rangle}=\\=
\Sum_{i=1}^n\left(\frac{\frac12\bs\theta_i^2}{z-t_i}+\frac{\d_i}{z-t_i}\log\langle\bs\theta_{\infty}| V_{\bs\theta_{n-2}}(t_{n-2}) \ldots
V_{\bs\theta_1}(t_1)|\bs\theta_0\rangle\right)\label{Tval}
}
where we have extended this formula to include $t_1=0$ and $t_n=\infty$.

Now solving the linear system \rf{lin_system} with $A(z) =\sum_i \frac{A_i}{z-t_i}$ we get
\eq{\Phi(z+\frac t2)\Phi(z-\frac t2)^{-1}=\Phi(z)\left(1+\frac t2 A(z)+\frac{t^2}8(\d A(z)+A(z)^2)+\ldots\right)\times\\\times
\left(1+\frac t2 A(z)+\frac{t^2}8(-\d A(z)+A(z)^2)+\ldots\right)\Phi(z)^{-1}=\\=
\Phi(z)\left(1+t A(z)+\frac{t^2}2A(z)^2+\ldots\right)\Phi(z)^{-1}
}
Therefore, due to the definition of the tau-function
\eq{\Tr\ \Phi(z+\frac t2)\Phi(z-\frac t2)^{-1}=\frac{t^2}2\Sum_{i=1}^n\left(\frac{\frac12\bs\theta_i^2}{(z-t_i)^2}+\frac{\d_i}{z-t_i}
\log\tau(t_1,\ldots,t_n)\right)+\ldots
}
Comparing this formula with \rf{Tval} completes the proof.
\hfill$\square$

\subsection{Fredholm determinant}

Consider now the isomonodromic tau-function $\tau(t) = \langle\bs\theta_\infty| V_{\bs\nu_1}(1)V_{\bs\nu_t}(t)|\bs\theta_0\rangle$, corresponding to the problem on sphere with four marked points
at $z=0,t,1,\infty$. Inserting the resolution of unity one can write
\be
\tau(t) = \langle\bs\theta_\infty| V_{\bs\nu_1}(1)V_{\bs\nu_t}(t)|\bs\theta_0\rangle =
\Sum_{\mathbf{Y},\mathbf{m}}
\langle\bs\theta_\infty|V_{\bs\nu_1}(1)|\mathbf{Y},\mathbf{m};\bs\sigma\rangle\langle
\mathbf{Y},\mathbf{m};\bs\sigma|V_{\bs\nu_t}(t)|\bs\theta_0\rangle =
\\
= \Sum_{\{\{p_{\alpha,i}\},\{q_{\alpha,i}\}\}}
\langle\bs\theta_\infty|V_{\bs\nu_1}(1)|\{p_{\alpha,i}\},\{q_{\alpha,i}\};\bs\sigma\rangle
\langle \{q_{\alpha,i}\},\{p_{\alpha,i}\};\bs\sigma|V_{\bs\nu_t}(t)|\bs\theta_0\rangle
\label{exptau}
\ee
Here we have used first just a particular case of the expansion \rf{expK}, applying it to the
simplest nontrivial isomonodromic tau-function. However, now it is useful to notice, that summation over the basis
in total space $\mc H^{\bs\sigma}=\bigoplus\limits_{\bs m\in\mathbb{Z}^N}\mathcal{H}^{\bs\sigma}_{\bs m}$
can be performed in Frobenius coordinates just forgetting restriction $\# p_\alpha=\#q_\alpha$ for the
states \rf{sigstates} in $\mathcal{H}^{\bs\sigma}_{\bs m}$, hence there is no restriction in summation range
in the r.h.s. of \rf{exptau}.

Now, one can still apply formulas \rf{maelK1}, \rf{maelK2} for the matrix elements in \rf{exptau}.
It gives
\be
\langle\bs\theta_\infty|V_{\bs\nu_1}(1)|\{p_{\alpha,i}\},\{q_{\alpha,i}\};\bs\sigma\rangle =
\det K_{x_Iy_J}
\\
\langle \{p_{\alpha,i}\},\{q_{\alpha,i}\};\bs\sigma|V_{\bs\nu_t}(t)|\bs\theta_0\rangle =
\det \tilde{K}_{x_Iy_J}(t)
\\
\tilde{K}_{p_\alpha,q_\beta}^{\alpha\beta}(t) = t^{p_\alpha+q_\beta-\sigma_\alpha+\sigma_\beta}
\tilde{K}_{p_\alpha,q_\beta}^{\alpha\beta}
\ee
where we have used again the multi-indices $\cup_\alpha\{(\alpha,p_{\alpha,i})\}=\{x_I\}$
and $\cup_\alpha\{(\alpha,q_{\alpha,i})\}=\{y_J\}$. It means, that the tau-function \rf{exptau}
can be summed up into a single Fredholm determinant
\eq{
\tau(t)
= \Sum_{\{\{p_{\alpha,i}\},\{q_{\alpha,i}\}\}}
\langle\bs\theta_\infty|V_{\bs\nu_1}(1)|\{p_{\alpha,i}\},\{q_{\alpha,i}\};\bs\sigma\rangle
\langle \{q_{\alpha,i}\},\{p_{\alpha,i}\};\bs\sigma|V_{\bs\nu_t}(t)|\bs\theta_0\rangle=
\\
=\Sum_{\{x\},\{y\}}\det K_{x,y}\cdot\det\tilde K_{y,x}(t) =
\Sum_{n=0}^\infty\Sum_{\genfrac{}{}{0pt}{}{|\{x\}|=n}{|\{y\}|=n}}\det K_{x,y}\cdot\det\tilde K_{y,x}(t)=
\\
=\Sum_{n=0}^\infty \Tr \wedge^n( K\tilde K(t))=\det(1+K\tilde K(t)) = \det\left(1+\mc{R}_t\right)
}
where basically only the Wick theorem has been used. One can also present the kernel of this operator $\mc{R}_t=K\tilde K(t)$ explicitly by the formula
\be
\mc{R}(x,z)=\Oint_{|y|=r}\frac{(\phi(x)\phi(y)^{-1}-x^{\bs\sigma}y^{-\bs\sigma})(S^{-1}\tilde\phi(y/t)\tilde\phi(z/t)^{-1}
S-y^{\bs\sigma}z^{-\bs\sigma})}{t^{-1}(x-y)(y-z)}dy
\ee
(where $S$ is the diagonal matrix introduced before),
so that this integral operator acts from the space of vector-valued functions $\bs f(z)=(f_1(z),\ldots,f_N(z))$ on the circle $|z|=r$, $t<r<1$. These functions have the fractional
Laurent expansion
\be
f_\alpha(z)=z^{\sigma_\alpha}\Sum_{n\in\mathbb Z}f_{\alpha,n}z^n
\ee
otherwise their convolution with our kernel will be ill-defined.

The representation in terms of the Fredholm determinant definitely requires further careful investigation,
and it could appear to be useful for practical computations with isomonodromic tau-functions, which basically
have no explicit representations.

\setcounter{equation}0
\section{Isomonodromic solutions to the Toda lattices
\label{ss:tautoda}}

\subsection{Matrix elements and tau-functions}

The multi-point analogs of the matrix elements \rf{mael}
\be
Z(\{\theta,t\}|Y',Y) = \langle\theta',Y'|\prod_kV_{\theta_k}(t_k)|Y,\theta\rangle
\ee
can be also computed using the Wick theorem. However, much simpler exercise is to compute their
generating functions, or the Toda lattice tau-functions
\be
\tau(\bs T,\bs{\bar T}|\{\theta,t\}) = \sum_{Y,Y'}Z(\{\theta,z\}|Y',Y)s_{Y'}(\bs T)s_Y(-\bs{\bar T})
=\langle\theta'|e^{H(\bs T)}\prod_k V_{\theta_k}(t_k) e^{-\bar{H}(\bs{\bar T})}|\theta\rangle
\ee
after introducing
\be
H(\bs T) = \sum_{k>0}T_kJ_k,\ \ \ \ \bar{H}(\bs{\bar T}) = \sum_{k>0}\bar{T}_kJ_{-k}
\ee
and using, that
\be
e^{-\bar{H}(\bs{\bar T})}|\theta\rangle = \sum_Y s_Y(-\bs{\bar T})|Y,\theta\rangle,\ \ \
\langle\theta'|e^{H(\bs T)} = \sum_Y s_Y(\bs T)\langle \theta',Y|
\ee
The result of immediate computation gives
\be
\tau(\bs T,\bs{\bar T}|\{\theta,t\}) =
\delta_{\theta',\theta+\sum_k\theta_k}\prod_k t_k^{\theta\theta_k}\prod_{i<j}(t_i-t_j)^{\theta_i\theta_j}\times
\\
\times
\exp\Sum_{n>0}\left(- n T_n\bar T_n+T_n\Sum_k \theta_k t_k^n+\bar T_n\Sum_k \theta_k t_k^{-n}\right)
\label{tautriv}
\ee
As a function of $T$-times this is just the vacuum tau-function of the Toda lattice hierarchy with $\{\theta,t\}$-dependent linear shift of times. However, as a function of $\{t\}$-variables this turns to be the simplest example of the tau-function of an isomonodromic deformation problem.

\subsection{Generalized integrable hierarchies}

From theorem~\ref{thm:bilinear_operator} it follows for any (radially ordered) product ${\mc O}=
\prod_j V_{\bs\nu_{j}}(t_j)$ of the monodromy vertex operators, that for any $k\in \mathbb{Z}$
\eq{
\langle\bs n,\bs\theta_\infty|e^{\Sum_{i>0,\alpha}T_{\alpha,i} J_{\alpha,i}}\otimes \langle\bs n',\bs\theta_\infty|e^{\Sum_{i>0,\alpha}
T'_{\alpha,i} J_{\alpha,i}}\cdot [\mc O\otimes\mc O,\mc I_k]\times\\\times
e^{-\Sum_{i<0,\alpha}\bar T_{\alpha,i} J_{\alpha,-i}}|\bar{\bs n},\bs\theta_0\rangle\otimes e^{-\Sum_{i<0,\alpha}\bar T'_{\alpha,i} J_{\alpha,-i}}
|\bar{\bs n}',\bs\theta_0\rangle=0
\label{obvious1}
}
Using \rf{bosonization}, \rf{Xmodes} and \rf{heisenberg} one can rewrite it as
{\small
\eq{\Sum_\alpha\epsilon_\alpha(\bs n+\bs n')\Oint_\infty dz z^{k+n_\alpha-n'_\alpha-2} e^{\xi_\alpha(\bs T-\bs T',z)}\cdot
\\
\cdot\tau(\bs n-\bs 1_\alpha,\bs T-[z^{-1}]_\alpha,\bar{\bs n},\bar{\bs T})
\tau(\bs n'+\bs 1_\alpha,\bs T'+[z^{-1}]_\alpha,\bar{\bs n}',\bar{\bs T}')=
\\
= \Sum_\alpha\epsilon_\alpha(\bar{\bs n}+\bar{\bs n}')\Oint_0 dzz^{k+\bar n_\alpha-\bar n'_\alpha}e^{\xi_\alpha(\bar{\bs T}-\bar{\bs T}',z^{-1})}\cdot
\\
\cdot\tau(\bs n,\bs T,\bar{\bs n}+\bs 1_\alpha,\bar{\bs T}-[z]_\alpha)
\tau(\bs n,\bs T,\bar{\bs n}-\bs 1_\alpha,\bar{\bs T}+[z]_\alpha)
\label{iToda}
}}
where
\eq{
\left[z\right]_\alpha=\,\bs 1_\alpha\otimes\left(1,z,\frac{z^2}{2},\frac{z^3}{3},\ldots\right)\,,\qquad
\left[z^{-1}\right]_\alpha=\,\bs 1_\alpha\otimes\left(1,\frac1{z},\frac1{2z^2},\frac1{3z^3},\ldots\right)
\\
\xi_\alpha(\bs T,x)= \Sum_{i=1}^\infty T_{\alpha,i}x^i
}
This is an \emph{infinite} collection of the bilinear equations for the tau-function, labeled
by integer number $k$, which contains the equation for $k=0$ corresponding to ordinary $N$-component two-dimensional Toda lattice (2DTL) hierarchy
(actually it is the same as $2N$-component KP hierarchy, as is clearly seen from \rf{iToda}).

It is especially interesting to obtain the system of bilinear equations only in the $\{\bs T,\bs n\}$ variables (they exist in closed form only for $k\geq0$, otherwise
there are unavoidable contributions from extra poles in the r.h.s. of \rf{iToda}). Substituting $\bar{\bs T}=\bar{\bs T}'$, $\bar{\bs n}=\bar{\bs n}'$ one obtains
\eq{\small
\Sum_\alpha\epsilon_\alpha(\bs n+\bs n')\Oint_\infty dz z^{k+n_\alpha-n'_\alpha-2} e^{\xi_\alpha(\bs T-\bs T',z)}\cdot
\\
\cdot\tau(\bs n-\bs 1_\alpha,\bs T-[z^{-1}]_\alpha,\bar{\bs n},\bar{\bs T})
\tau(\bs n'+\bs 1_\alpha,\bs T'+[z^{-1}]_\alpha,\bar{\bs n},\bar{\bs T})=0
\label{iKP}
}
where $\{\bar{\bs T},\bar{\bs n}\}$ now play the role of parameters of the solutions (this hierarchy can be called as $N$-component isomonodromic KP, in contrast to
ordinary case  $N$-component isomonodromic 2DTL is not equivalent to  $2N$-component isomonodromic KP).
We are going to return to these hierarchies in detail elsewhere, and now let us present
just few examples.

\subsection{Examples of isomonodromic hierarchies}

The simplest example is the single-component isomonodromic 2DTL, given by the following bilinear equations
\eq{\label{1-toda}
\oint_\infty dz z^{k-n+n'-2} e^{\xi(\bs T-\bs T',z)}\tau(n+1,\bs T-[z^{-1}],\bar n,\bar{\bs T})
\tau(n'-1,\bs T'+[z^{-1}],\bar n',\bar{\bs T'})=
\\
=
\oint_0 dz z^{k-\bar n+\bar n'}e^{\xi(\bar{\bs T}-\bar{\bs T}',z^{-1})}\tau(n,\bs T,\bar n-1,\bar{\bs T}-[z])\tau(n',\bs T,\bar n'+1,\bar{\bs T}+[z])
}
Since all quasi-group operators have definite charges, here one can use simple parametrization
$n=\bar n+q-1$, $n'=\bar n'+q+1$, and introduce
\eq{\tau_n(\bs T,\bar{\bs T})=\tau(n+q,\bs T,n,\bar{\bs T})
}
so that equations \rf{1-toda} turn into
\eq{
\oint_\infty dz z^{k-n+n'} e^{\xi(\bs T-\bs T',z)}\tau_{n}(\bs T-[z^{-1}],\bar{\bs T})
\tau_{n'}(\bs T'+[z^{-1}],\bar{\bs T'})=\\=
\oint_0 dz z^{k-n+n'}e^{\xi(\bar{\bs T}-\bar{\bs T}',z^{-1})}\tau_{n-1}(\bs T,\bar{\bs T}-[z])\tau_{n'+1}(\bs T',\bar{\bs T}'+[z])
}
It is easy to show, that this hierarchy has only the shifted vacuum solution
\eq{
\tau_n(\bs T,\bar{\bs T})=Ae^{b\, n}\exp\Sum_{l>0}\left(-lT_l\bar T_l+\gamma_l T_l+\delta_l\bar T_l\right)
}
Substituting here $\gamma=\Sum_i\theta_i[t_i]$, $\delta=\Sum_i\theta_i[t_i^{-1}]$ one gets
exactly the $\{T,\bar{T}\}$-times dependent part of \rf{tautriv}.

As a next example consider the isomonodromic two-component KP, which is already a non-trivial integrable system. The bilinear equations now are
\eq{
\oint_\infty dz z^{k-n_1+n'_1-2}e^{\xi_1(\bs T-\bs T',z)}\tau(n_1+1,n_2,\bs T-[z^{-1}]_1)
\tau(n_1'-1,n_2',\bs T'+[z^{-1}]_1)+\\+
(-1)^{n_1+n_1'}\oint_\infty dz z^{k-n_2+n'_2-2}e^{\xi_2(\bs T-\bs T',z)}\cdot
\\
\cdot\tau(n_1,n_2+1,\bs T-[z^{-1}]_2)\tau(n_1',n_2'-1,\bs T'+[z^{-1}]_2) = 0
}
Now one has fixed total charge $n_1+n_2=q$, so we parameterize $n_1=n-1, n_2=q-n, n_1'=n'+1, n_2'=q-n_2$
and get
\eq{
\oint_\infty dz z^{k-n+n'}e^{\xi_1(\bs T-\bs T',z)}\tau_{n}(\bs T-[z^{-1}]_1)\tau_{n'}(\bs T'+[z^{-1}]_1)+\\+
(-1)^{n+n'}\oint_\infty dz z^{k+n-n'-2}e^{\xi_2(\bs T-\bs T',z)}\tau_{n-1}(\bs T-[z^{-1}]_2)\tau_{n+1}(\bs T'+[z^{-1}]_2)=0
\label{isotoda1}
}
and the sign factor can be simplified by redefinition $\tau_n(\bs T)\mapsto (-1)^{\frac{n^2+\chi(n)}2}\tau_n(\bs T)$, where $\chi(n)$ is a mod 4 Dirichlet character:
$\chi(0)=\chi(2)=0$, $\chi(1)=1$, $\chi(3)=-1$. One gets therefore
\eq{
\oint_\infty dz z^{k-n+n'}e^{\xi_1(\bs T-\bs T',z)}\tau_{n}(\bs T-[z^{-1}]_1)\tau_{n'}(\bs T'+[z^{-1}]_1)=\\=
\oint_0 dz z^{-k-n+n'}e^{\xi_2(\bs T-\bs T',z^{-1})}\tau_{n-1}(\bs T-[z]_2)\tau_{n+1}(\bs T'+[z]_2)
}
where in contrast to \rf{iToda} the equation label $k$ enters two integrals with different signs.
We present finally first several equations of this hierarchy:
\begin{itemize}
\item $k=0$, $n=n'$:\par $D_{1,1}D_{2,1}\tau_n\cdot\tau_n+2\tau_{n+1}\cdot\tau_{n-1}=0$,\par $D_{1,2}D_{2,1}\tau_n\cdot\tau_n=2D_{1,1}\tau_{n+1}\cdot\tau_{n-1}$,\par
$(D_{1,1}^4+3D_{1,1}^2-4D_{1,1}D_{1,3})\tau_n\cdot\tau_n=0$,\par
\ldots
\item $k=1$, $n=n'$:\par $D_{1,1}^2\tau\cdot\tau=2\tau_{n+1}\cdot\tau_{n-1}$,\par $D_{1,2}D_{2,1}\tau\cdot\tau+2D_{2,1}\tau_{n+1}\cdot\tau_{n-1}=0$,\par $D_{1,2}^2\tau_n\cdot\tau_n=2D_{1,1}^2\tau_{n+1}\cdot\tau_{n-1}$,\par
$(D_{1,1}^4-3D_{1,2}^2+8D_{1,1}D_{1,3})\tau_n\cdot\tau_n+12D_{1,2}\tau_{n+1}\cdot\tau_{n-1}=0$,\par\ldots
\end{itemize}

\section{Conclusion}

We have considered in this paper the free fermion formalism, which allows to study representations
of the W-algebras at least at integer values of the central charges. The vertex operators are defined by
their two-fermion matrix elements, which are fixed by monodromies of auxiliary linear system, and
can be obtained from solution of the corresponding Riemann-Hilbert problem.

This paper is just the first step of studying this relation (apart of the well-known and effectively used
for different applications Abelian case). A natural development of the above ideas is only outlined
in sect.~\ref{ss:tauiso} and \ref{ss:tautoda}. We are going to return elsewhere to the problem of
rewriting the isomonodromic tau-functions in terms of the Fredholm determinants, which can be quite
useful representations (though still not an explicit form) for these complicated objects. Another
point, which has to be understood better is the relation of class of the isomonodromic solutions to
the Toda lattices, which have been defined above using the generalized Hirota bilinear relations, to
the class of solutions, obeying the Virasoro-W constraints.

Finally, it would be extremely interesting to study the relation of generic W-conformal blocks and
isomonodromic tau-functions to the topological strings -- at least on the level of topological vertices,
and to the mostly intriguing four-dimensional (supersymmetric) quantum gauge theories -- the main subject of interests of Igor Tyutin.

\subsection*{Acknowledgements}

The work was supported by the joint Ukrainian-Russian (NASU-RFBR) project 01-01-14 (P.G.) and 14-01-90405 (A.M.), the work of P.G. has been also supported by joint NASU-CNRS project F14-2015 and by the ``Young Russian Mathematics'' stipend, while the work of A.M. was also supported by RFBR  grant 15-01-99504, by joint RFBR/JSPS project 15-51-50034 and by the Program of Support of Scientific Schools (NSh-1500.2014.2).

This paper was also prepared within the framework of a subsidy granted to the National Research University Higher School of Economics by the Government of the Russian Federation for the implementation of the Global Competitiveness Program.


\begin{thebibliography}{10}

\bibitem{ZamW}
A.~Zamolodchikov,
Theor. Math. Phys, {\bf 65}:3, (1985), 1205–1213.


\bibitem{FZ}V.~Fateev and A.~Zamolodchikov,
Nucl. Phys. {\bf B280}, (1987), 644-660.

\bibitem{FL}V.~Fateev and S.~Lukyanov,
Int. J. Mod. Phys. {\bf A3} (2), (1988), 507-520.

\bibitem{Wblocks}P.~Bowcock and G.M.T.~Watts,
Theor.~Math.~Phys. {\bf 98}, (1994), 350-356
[arXiv:hep-th/9309146].

\bibitem{BPZ} A.~Belavin, A.~Polyakov and A.~Zamolodchikov,
Nucl. Phys. {\bf B241}, (1984), 333-380.

\bibitem{LMN}
A.~S.~Losev, A.~Marshakov and N.~Nekrasov, in Ian Kogan memorial volume
{\it From fields to strings: circumnavigating theoretical physics},
581-621; [hep-th/0302191].

\bibitem{NO}
N.~Nekrasov and A.~Okounkov,
[hep-th/0306238].

\bibitem{Wiegmann}E.~Bettelheim, A.~Abanov, P.~Wiegmann,
J.Phys. A40 (2007) F193-F208, [arXiv:nlin/0605006 [nlin.SI]].

\bibitem{AGT}
 L.~F.~Alday, D.~Gaiotto and Y.~Tachikawa,
  Lett.\ Math.\ Phys.\  {\bf 91} (2010) 167
  [arXiv:0906.3219 [hep-th]].

\bibitem{Nek}
N.~Nekrasov,
  Adv.\ Theor.\ Math.\ Phys.\  {\bf 7} (2004) 831
  [arXiv:hep-th/0206161];\\
 N.~Nekrasov and V.~Pestun,
  [arXiv:1211.2240 [hep-th]].


\bibitem{SW}N.~Seiberg and E.~Witten
Nucl. Phys. {\bf B426}, 19, (1994),
[arXiv:hep-th/9407087].

\bibitem{GMtw}P.~Gavrylenko, A.~Marshakov,
JHEP {\bf 0216}, (2016), 181,
[arXiv:1507.08794 [hep-th]].

\bibitem{ZamAT}
Al.~Zamolodchikov, Nucl. Phys. \textbf{B285}, [FS19], (1987), 481-503; JETP {\bf 90}, (1986), 1808-1818.\\
S.~Apikyan and Al.~Zamolodchikov,
JETP {\bf 92}, (1987), 34-45.

\bibitem{KriW}
I.~Krichever,
Commun. Pure. Appl. Math. {\bf 47} (1994) 437
[arXiv: hep-th/9205110].

\bibitem{AMtau}
  A.~Marshakov,
  JHEP {\bf 1307} (2013) 068,
  [arXiv:1303.0753 [hep-th]].

\bibitem{Quiver}P.~Gavrylenko and A.~Marshakov,
JHEP {\bf 0514}, (2014), 097,  [arXiv:1312.6382 [hep-th]].

\bibitem{Painleve}
  O.~Gamayun, N.~Iorgov and O.~Lisovyy,
  JHEP,  {\bf 1210}, (2012), 38,
  [hep-th/1207.0787].

\bibitem{PG}
P.~Gavrylenko ,
JHEP {\bf 0915}, (2015), 167,
[arXiv:1505.00259 [hep-th]].


\bibitem{Litv1}
V.~Fateev and A.~Litvinov,
JHEP {\bf 0711}, (2007), 002,
[arXiv:0709.3806 [hep-th]].

\bibitem{SJM} M.~Sato, T.~Miwa and M.~Jimbo,
Publ. RIMS Kyoto Univ. {\bf 14},
(1978), 223–267; {\bf 15}, (1979), 201–278; {\bf 15}, (1979), 577–629;{\bf 15},
(1979), 871–972; {\bf 16}, (1980),
531–584.

\bibitem{NSlav}N. Kitanine, K. K. Kozlowski, J. M. Maillet, N. A. Slavnov, V. Terras,
J. Stat. Mech. (2011) P12010,
[arXiv:1110.0803 [hep-th]].

\bibitem{Kozlowski}K.K.~Kozlowski,
[arXiv:1501.07711 [math-ph]].

\bibitem{Pogr} A.~Pogrebkov,
Russian Mathematical Surveys (2003), 58(5):1003

\bibitem{OP}
A.~Okounkov and R.~Pandharipande, \emph{Gromov-Witten theory, Hurwitz theory, and completed cycles},
arXiv:math.AG/0204305; \emph{The equivariant Gromov-Witten theory of} $\mathbb{P}^1$, arXiv:math.AG/0207233.

\bibitem{MN}
A.~Marshakov and N.~Nekrasov,
JHEP {\bf 0701} (2007) 104,
[arXiv: hep-th/0612019].

\bibitem{Litv3}V.A.~Fateev, A.V.~Litvinov,
JHEP {\bf 1201}, (2012), 051,
[arXiv:1109.4042 [hep-th]].

\bibitem{ILT} N.~Iorgov, O.~Lisovyy, J.~Teschner,
Comm.~Math.~Phys. {\bf 336}, (2015), 671-694
[arXiv:1401.6104~[hep-th]].

\bibitem{PoPya}A.~Marshakov,
Theor.Math.Phys.154:362-384, (2008), [arXiv:0706.2857 [hep-th]].

\bibitem{Hirota}T.~Miwa, M.~Jimbo, E.~Date,
{\sl Solitons: Differential Equations, Symmetries and Infinite Dimensional Algebras,}
Cambridge University Press, 2000.\par
V.G. Kac, J.W. van de Leur,
J.Math.Phys. 44 (2003) 3245-3293,
[arXiv:hep-th/9308137].


\bibitem{AZ}
A.~Alexandrov and A.~Zabrodin,
J.Geom.Phys. 67 (2013) 37-80,
[arXiv:1212.6049 [math-ph]].


\end{thebibliography}
\end{document}